\pdfoutput=1

\documentclass{article}

\usepackage{microtype}
\usepackage{graphicx}
\usepackage{subcaption}
\usepackage{booktabs} %
\usepackage{times} 

\usepackage{hyperref}

\usepackage[accepted]{icml2026}

\usepackage{calc}
\usepackage[dvipsnames,table]{xcolor}
\usepackage[export]{adjustbox}
\usepackage{mathtools}
\usepackage{pifont}
\usepackage[subtle]{savetrees}
\usepackage{multirow}
\usepackage{makecell}
\usepackage{extarrows}
\usepackage{scrextend}
\usepackage{adjustbox}
\usepackage{algorithm}
\makeatletter
\@ifundefined{algorithmic}{}{%

}
\makeatother
\usepackage{algpseudocode}
\usepackage{amsmath}
\usepackage{amssymb}
\usepackage{amsthm}

\usepackage[capitalize,noabbrev]{cleveref}

\theoremstyle{plain}
\newtheorem{theorem}{Theorem}[section]

\newtheorem{lemma}[theorem]{Lemma}

\theoremstyle{definition}
\newtheorem{definition}[theorem]{Definition}
\newtheorem{assumption}[theorem]{Assumption}
\theoremstyle{remark}

\definecolor{best}{gray}{0.85}      %
\definecolor{second}{gray}{0.93}    %
\usepackage[textsize=tiny]{todonotes}

\newcommand{\projectpageurl}{https://yingheng-wang.com/DeepSciReasoner}
\newcommand{\codeurl}{https://github.com/yingheng-wang/DeepSciReasoner}
\definecolor{projectlink}{RGB}{200,93,49}
\newcommand{\projecturl}[1]{{\hypersetup{urlcolor=projectlink}\href{#1}{#1}}}

\icmltitlerunning{DeepSciReasoner}

\begin{document}

\twocolumn[
  \icmltitle{Deep Scientific Reasoning under Physical Constraints: \\Structure-Aware Spectrum Prediction}

  \begin{icmlauthorlist}
    \icmlauthor{Yingheng Wang}{yyy}
    \icmlauthor{Tao Yu}{zzz}
    \icmlauthor{Shufeng Kong}{yyy}
    \icmlauthor{Francesco Ricci}{fr1,fr2}
    \icmlauthor{John M. Gregoire}{jg1,jg2}
    \icmlauthor{Carla P. Gomes}{yyy}
  \end{icmlauthorlist}

  \icmlaffiliation{yyy}{Department of Computer Science, Cornell University}
  \icmlaffiliation{jg1}{California Institute of Technology}
  \icmlaffiliation{jg2}{Lila Sciences}

  \icmlaffiliation{zzz}{Griffith University}

  \icmlaffiliation{fr1}{Institute of Condensed Matter and Nanosciences, Université catholique de Louvain}

  \icmlaffiliation{fr2}{Matgenix SRL}

  \icmlcorrespondingauthor{Yingheng Wang}{yingheng@cs.cornell.edu}

  \icmlkeywords{Scientific Reasoning, Spectrum Prediction}

  \vskip 0.3in
]

\makeatletter
{\let\dsr@notice\Notice@String
 \renewcommand{\Notice@String}{Code: \projecturl{\codeurl}\\\dsr@notice}%
 \printAffiliationsAndNotice{}}  %
\makeatother

\begin{abstract}
    Structured scientific spectra encode rich physical information while obeying hard constraints, such as conservation laws and characteristic spectral geometry. Accurate prediction of these spectra is central to materials discovery, yet existing methods often treat them as unconstrained sequences and therefore fail to enforce the underlying physical structure. Taking electronic density of states (eDOS) as the prototypical example, we introduce \textbf{DeepSciReasoner}, a general paradigm for predicting scientific spectra under physical constraints. The framework combines structure-aware spectrum decoding with constraint-preserving physical reasoning, allowing predictions to capture rich spectral structure while respecting the underlying physics. We evaluate DeepSciReasoner on eDOS, phonon density of states (phDOS), X-ray absorption near-edge structure (XANES), and Raman spectra, where it substantially improves prediction accuracy while maintaining physical consistency. These results establish DeepSciReasoner as a reusable blueprint for structured scientific spectrum prediction under hard physical constraints.
Code and data are available on the project page:\par\vspace{-\parskip}
{\centering\projecturl{\projectpageurl}\par}
\end{abstract}

\section{Introduction}

Structured scientific spectra arise throughout the physical sciences, including electronic density of states (eDOS), phonon density of states (phDOS), X-ray absorption near-edge structure (XANES), Raman spectra, mass spectra, and nuclear magnetic resonance spectra. These spectra encode rich mechanistic information while obeying hard physical constraints. These constraints include conservation laws (e.g., mass/charge/state counts) and spectral geometry (e.g., sharp discontinuities at band edges and peaks) that are essential for correct downstream observables. 

Machine learning has made rapid progress on crystal property prediction for scalar targets~\citep{Xie2018,schutt2018schnet,Chen2019}. Predicting physical spectra is more challenging because the output is a high-dimensional sequence over a spectral axis (energy, wavenumber, or frequency) whose validity depends on physical constraints and spectral geometry. Beyond basic nonnegativity, physically meaningful predictions must preserve the \emph{total spectral mass} (the integral over the spectral axis) and accurately capture \emph{where spectral mass lies}, including sharp band edges, prominent peaks, and sparse vibrational lines. Taking eDOS as the prototypical example, total-state conservation, band-edge sharpness, and the placement of Van Hove singularities~\citep{van1953occurrence} all need to be respected for the prediction to be physically faithful; analogous geometric and conservation considerations recur in phDOS, XANES, and Raman.

Existing approaches~\citep{chen2021direct,kaundinya2022prediction,Kong2022,lee2023density} often fall short, revealing a mismatch between pointwise regression and the constrained geometric nature of physical spectra. For example, there are three key limitations in previous eDOS prediction methods. First, they predict the band gap and spectrum separately and reconcile them post hoc, which misaligns band edges near the Fermi level and introduces spurious in-gap density. Second, pointwise regression loss can report low per-bin error while still shifting peaks, blurring features such as Van Hove singularities, or drifting from global constraints, yielding spectra that are visually plausible yet physically inconsistent and that induce erroneous derived properties. Third, the total eDOS is theoretically a sum of atom-projected DOS (pDOS)~\citep{mulliken1955electronic}; however, prior models collapse a crystal into a single global embedding, reducing interpretability and ignoring atom-projected contribution. These pathologies: mass drift, peak blurring, and missing element-resolved attribution, are not unique to eDOS and recur across other structured spectra.

These issues motivate moving beyond unconstrained regression to reasoning over structured spectra under physical constraints. To this end, we introduce \textbf{DeepSciReasoner}, a general spectrum prediction framework that couples structure-aware spectrum prediction with constraint-preserving physical reasoning. This framework generalizes across structured spectra with the same architecture essentially unchanged.

DeepSciReasoner has two stages: (i) an edge-aware crystal graph Transformer encodes atomic structure with gated, edge-modulated attention; and a structure-aware spectrum decoder operates over discrete spectral bins using source (atom-to-spectral-axis) attention, preserving physically meaningful atom-projected contributions~\citep{vaswani2017attention}; (ii) a mass-conserving iterative refinement transports the normalized spectral density with a learned velocity field that satisfies the continuity equation~\citep{kundu2024fluid}. The velocity is supervised using a 2-Wasserstein displacement-interpolation reference flow~\citep{mccann1997convexity}, and inference is implemented stably via a CDF-based upwind scheme~\citep{godunov1959finite}. Together, this method produces spectra with sharper edges and peaks while preserving total spectral mass.

\vspace{-8pt}
\paragraph{Contributions}
In summary, we make the following contributions. (i) We introduce DeepSciReasoner, a general design paradigm for deep scientific reasoning under hard physical constraints for structure-aware spectrum prediction, formulated generically over discrete spectral grids so that the same architecture transfers across spectrum types. (ii) Using eDOS as the prototypical example, we introduce a crystal-to-spectrum model with source attention from atoms to spectral bins, designed to retain element-resolved contributions across the spectrum, and (iii) a mass-conserving iterative refinement stage that transports the normalized spectral density with a continuity-constrained velocity field to sharpen peaks and gaps while preserving total spectral mass. (iv) Across four physically distinct structured spectra (eDOS, phDOS, XANES, and Raman), DeepSciReasoner achieves state-of-the-art performance, demonstrating DeepSciReasoner as an extendable blueprint for reasoning over structured scientific spectra under physical constraints. (v) Beyond overall benchmarking, we show that DeepSciReasoner improves spectral gap identification, and exhibits stronger data-scaling behavior, demonstrating its practical value for physically constrained spectrum prediction.

\section{Related Work}

\paragraph{Scientific Reasoning under Constraints}
AI for scientific discovery~\citep{wang2023scientific} has advanced rapidly in recent years, with AlphaFold~\citep{jumper2021highly} standing out as a landmark: rather than a purely black-box predictor, it embeds physical constraints (e.g., triangle updates and rigid-body symmetry) that enable systematic reasoning about protein structure from sequence and evolutionary context. More broadly, AI has long been used for scientific reasoning. Early expert systems~\citep{lindsay1980applications} such as DENDRAL combined mass spectrometry with chemistry knowledge bases to infer candidate molecular structures~\citep{buchanan1981dendral}. Other examples include inverse-folding protein design~\citep{vucinic2020positive}, pedigree inconsistency detection~\citep{sanchez2008mendelian,o1999optimal}, and DNA word design~\citep{codish2017dna,frutos1997demonstration}. Recent work increasingly couples deep learning with principled reasoning under uncertainty such as crystal phase mapping~\citep{chen2021automatingdrnet} and multi-species population modeling~\citep{chen2016deep}, and more recently extends these ideas to large language models via reinforcement learning~\citep{wang2025scireasoner}. Motivated by this line, we study how deep learning and physical reasoning can be combined for structure-aware spectrum prediction.

\paragraph{Structured Spectrum Prediction}
Beyond scalar property, a growing line of work learns functional spectra directly from crystal structure, including the electronic density of states (eDOS), phonon density of states (phDOS), X-ray absorption near-edge structure (XANES), Raman, and beyond. Early efforts~\citep{fung2022physically,kaundinya2022prediction} cast spectrum prediction as a pointwise regression task using crystal representations alone. Recently, ``crystal-to-sequence'' formulations~\citep{lee2023density} have become common, treating the spectrum as an ordered sequence over a spectral axis and decoding it with sequential models. Related work extends to phonon DOS~\citep{chen2021direct,al2025rapid}, X-ray absorption~\citep{carbone2020machine,kotobi2023integrating,lin2025universal,mathew2018xas}, Raman~\citep{bagheri2023raman}, mass spectrum (MS)~\citep{nguyen2024advancing,goldman2024generating,youngfragnnet,wang2025neural}, and nuclear magnetic resonance (NMR)~\citep{cobas2020nmr,yang2021predicting}. In contrast to prior work, we explicitly enforce physical constraints in the learning algorithm: we treat each spectrum as an unnormalized density, decode it with structure-aware atom-to-spectral-axis attention, and refine it via mass-conserving transport to sharpen spectral geometry without breaking conservation. We present the framework with eDOS as the running example, and the same recipe transfers from eDOS to XANES, Raman, and phDOS without architectural changes.

\section{Method} \label{sec:method}
\begin{figure*}[ht]
    \centering
    \includegraphics[width=1\linewidth]{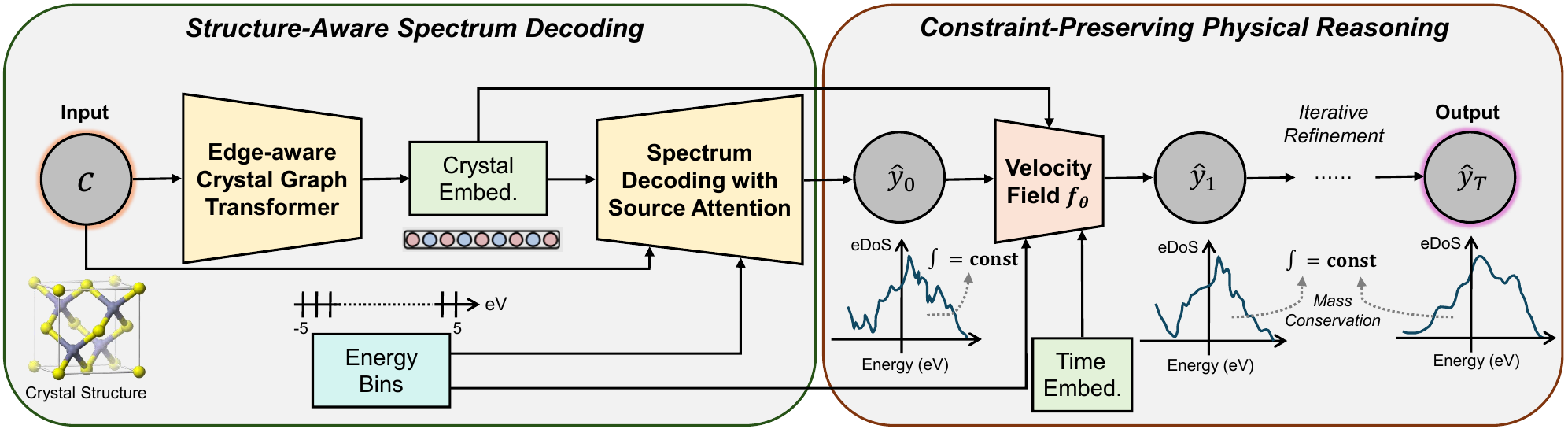}
    \caption{\textbf{DeepSciReasoner} consists of two stages, illustrated here with eDOS as the running example. (\emph{Left}) Structure-Aware Spectrum Decoding: the input crystal structure is encoded by an edge-aware crystal graph Transformer to produce atom-level embeddings, which are decoded into an initial spectrum via source (atom-to-spectral-axis) attention over discrete bins, preserving atom-projected contributions. (\emph{Right}) Constraint-Preserving Physical Reasoning: the initial normalized spectrum is refined through a learned velocity field that transports spectral mass along the axis while conserving total mass, yielding progressively sharper band edges and peaks and producing the final prediction.} \label{fig:main}
    \vspace{-10pt}
\end{figure*}

In this section, we introduce DeepSciReasoner, a physics-informed framework for structure-aware spectrum prediction. It proceeds in two stages (illustrated in \Cref{fig:main}): (i) structure-aware spectrum decoding, which embeds physical structure into the model for spectrum prediction, and (ii) constraint-preserving physical reasoning, which progressively improves the initial prediction while enforcing physical constraints. The formulation is general and the only target-specific choices are the spectral axis, its discretization, and optional conditioning. Throughout this section we take eDOS as the prototypical example, where the design explicitly captures atom-projected contributions across energies and preserves total-state conservation.

\vspace{-5pt}
\subsection{Preliminaries}\label{sec:pre}
\vspace{-5pt}
\paragraph{Problem Formulation}
We represent an input crystal $c$ as a graph $G_c = (V, \mathcal{E})$, where
$V = \{v_i\}_{i=1}^n$ is the set of $n$ atomic nodes and each node
$v_i \in \mathbb{R}^{d_v}$ denotes an atom embedding (e.g., element type and other attributes).
The edge set $\mathcal{E} = \{(i,j)\}$ encodes interatomic relations; for each edge
$(i,j) \in \mathcal{E}$, we associate an edge feature
$r_{ij} \in \mathbb{R}$ (or $r_{ij} \in \mathbb{R}^{d_e}$) representing the interatomic distance
(and optionally additional geometric information).
In crystalline systems, atomic arrangements repeat periodically in three-dimensional space;
we construct $\mathcal{E}$ using a local neighborhood graph by limiting each atom to at most
$n_{\max}$ neighbors for computational efficiency, so that the node degree is bounded by $n_{\max}$.
Given a crystal $c$, our goal is to predict a target spectrum
$y \in \mathbb{R}^{l_y}$ ($l_y$ spectral bins; e.g., eDOS, phDOS, XANES, or Raman) defined on a fixed discretized grid over a spectral axis $\epsilon$ (energy for eDOS/XANES, wavenumber/frequency for Raman/phDOS).
The learning problem is to approximate a mapping $f_\theta : G_c \;\mapsto\; y$.

\vspace{-5pt}
\paragraph{Physical Constraints of Structured Spectra}
A structured spectrum can be viewed as a (possibly unnormalized) density over its spectral axis $E\in[a,b]$. Taking eDOS as our running example, let $D(E)\ge 0$ denote the electronic density of states and define its normalized form as
$p(E) := \frac{D(E)}{\int_a^b D(u)\,du}.$
Physical validity requires \emph{nonnegativity}, $p(E)\ge 0$ for all $E\in[a,b]$, and \emph{total-mass conservation},
$\int_a^b p(E)\,dE = 1$ (for eDOS this is equivalent to $\int_a^b D(E)\,dE = N_{\text{states}}$); these constraints must be preserved throughout the reasoning trajectory (e.g., during iterative refinement).
Additionally, structured spectral features like sharp band edges/gaps and Van Hove singularities (and, analogously, white-line/post-edge features in XANES or sparse vibrational peaks in Raman/phDOS) need to be preserved.
This motivates formulating spectrum prediction as a physical reasoning problem rather than unconstrained pointwise regression. We provide more details about eDOS, phDOS, XANES, and Raman spectra in \Cref{app:edos}.

\vspace{-5pt}
\subsection{Structure-Aware Spectrum Decoding}\label{sec:seq}
\begin{figure}[ht]
\vspace{-10pt}
    \centering
    \includegraphics[width=1\linewidth]{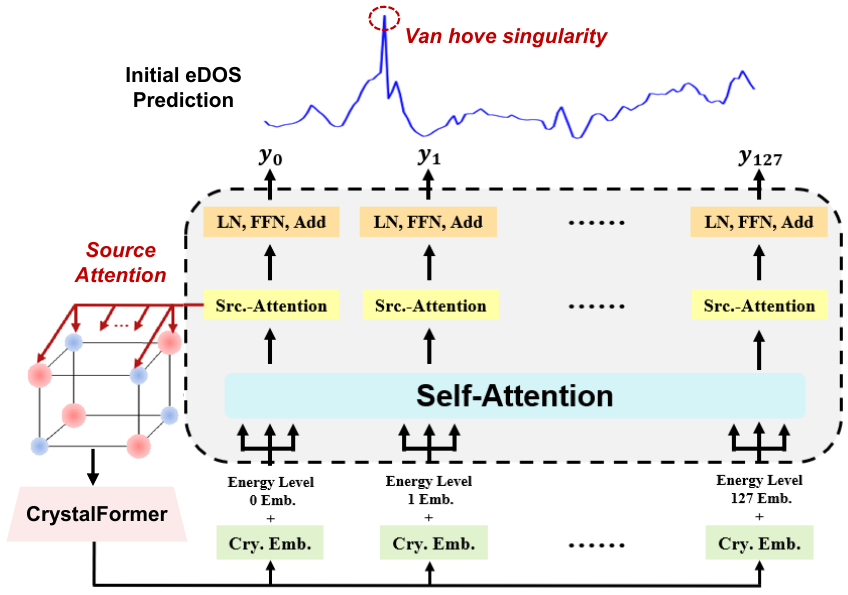}
    \caption{\textbf{Structure-aware Spectrum decoding.} The edge-aware crystal graph Transformer encodes the input structure and a decoder with hierarchical attention predicts the eDOS, where each energy bin selectively aggregates atom-projected contributions.} \label{fig:dec}
    \vspace{-10pt}
\end{figure}

We introduce a structure-aware atom-to-energy spectrum decoding framework with hierarchical attention (\Cref{fig:dec}): (i) an edge-aware crystal-graph Transformer encodes atomic structure via gated, edge-modulated attention, and (ii) a structure-aware decoder predicts over energy bins with source (atom-to-energy) attention, preserving physically meaningful atom-projected contributions.

\vspace{-5pt}
\paragraph{Structure Encoding with Edge-aware Attention}
We encode crystal structures using a crystal graph Transformer~\citep{velivckovic2017graph} with edge-aware attention~\citep{vaswani2017attention}, explicitly modeling both atomic and interatomic (edge) features.

Let $h_i^l \in \mathbb{R}^{d_h}$ denote the hidden representation of node $i$ in layer $l$, and $h_{e_{ij}} \in \mathbb{R}^{d_e}$ represent the embedding of edge feature between nodes $i$ and $j$. For each layer, we first project the node embeddings into query, key, and value spaces through learnable linear transformations:
\begin{equation}
    q_i^l = W_q^l h_i^l + b_q^l, \quad
    k_i^l = W_k^l h_i^l + b_k^l, \quad
    v_i^l = W_v^l h_i^l + b_v^l,
\end{equation}
where $W_q^l, W_k^l, W_v^l \in \mathbb{R}^{d \times d_h}$ and $b_q^l, b_k^l, b_v^l \in \mathbb{R}^{d}$ are layer-specific learnable parameters. 

Similarly, the edge embedding is linearly transformed into the same latent space: $g_{ij} = W_e h_{e_{ij}} + b_e$,
where $W_e \in \mathbb{R}^{d \times d_e}$ and $b_e \in \mathbb{R}^{d}$ are shared across all layers. This shared edge projection ensures consistent encoding of geometric or chemical bond information throughout the network.

The normalized attention coefficient between node $i$ and its neighbor $j$ is then computed as
\begin{equation}
    \alpha_{ij}^l = 
    \frac{\exp\left(\frac{(q_i^l)^\top (k_j^l + g_{ij})}{\sqrt{d}}\right)}
    {\sum_{p \in N(i)} \exp\left(\frac{(q_i^l)^\top (k_p^l + g_{ip})}{\sqrt{d}}\right)},
\end{equation}
where $N(i)$ denotes the set of neighboring nodes of $i$, and $\sqrt{d}$ serves as the standard scaling factor to stabilize training. 

The aggregated representation of node $i$ is computed as a weighted sum of its neighbors’ values, modulated by both attention scores and edge embeddings:
\begin{equation}
    \hat{h}_i^{l+1} = \sum_{j \in N(i)} \alpha_{ij}^l \cdot (v_j^l + g_{ij}).
\label{eq:gat}
\end{equation}
Since $v_j^l$ and $g_{ij}$ are projected into the same $d$-dimensional latent space, the additive modulation is well defined.
Compared to graph convolution~\citep{kipf2016semi,Xie2018}, this explicitly incorporates both pairwise similarity and interatomic features in message aggregation, thereby enhancing its expressiveness.

To improve training stability and support deeper stacks, we augment each layer with a \emph{gated residual} connection~\citep{he2016deep}. Let $\hat{h}_i^{l+1}$ denote the pre-residual output of layer $l$ (i.e., the representation after message passing/attention but before the residual update), and let $h_i^{l}$ be the input representation. We first project the residual branch to match dimensions, $r_i^{l} = W_r^{l} h_i^{l} + b_r^{l}$ with $W_r^{l}\in\mathbb{R}^{d\times d_h}$ and $b_r^{l}\in\mathbb{R}^{d}$,
and then compute a gate that adaptively interpolates between the new update $\hat{h}_i^{l+1}$ and the residual signal $r_i^{l}$. Specifically, we set
\begin{equation}
\begin{aligned}
\beta_i^{l} &= \sigma\!\left(a^{l}\,[\, \hat{h}_i^{l+1} ~;~ r_i^{l} ~;~ \hat{h}_i^{l+1} - r_i^{l}\,]\right), \\
h_i^{l+1} &= \sigma\!\Big(f_{\mathrm{LN}}\!\big((1-\beta_i^{l}) \odot \hat{h}_i^{l+1} + \beta_i^{l} \odot r_i^{l}\big)\Big),
\end{aligned}
\end{equation}
where $\sigma(\cdot)$ is the sigmoid in the gate and the nonlinearity after normalization, $[\cdot;\cdot]$ denotes concatenation, $f_{\mathrm{LN}}$ is LayerNorm, and $\odot$ is element-wise multiplication. The learnable gating matrix $a^{l}\in\mathbb{R}^{d\times 3d}$ maps the concatenated $3d$-dimensional vector to a per-channel gate $\beta_i^{l}\in\mathbb{R}^{d}$; for the final layer we omit the post-normalization $\sigma$ and $f_{\mathrm{LN}}$ so that the output embeddings $\{h_i^{L}\}$ are not squashed. When $\beta_i^{l}$ is close to $0$, the layer trusts the newly computed update $\hat{h}_i^{l+1}$; when it is close to $1$, the model falls back to the residual branch, improving robustness early in training.

This edge-aware, gated attention design enables expressive modeling of the crystal structure by dynamically weighting interatomic interactions and adaptively controlling information flow across layers, while preserving physically meaningful atom-projected contributions.

\vspace{-5pt}
\paragraph{Spectrum Decoding with Source Attention}
Existing approaches~\citep{Kong2022,kaundinya2022prediction,lee2023density} typically compress a crystal into a single global embedding and decode the eDOS with an MLP, which can obscure the atom-projected contributions that physically constitute the spectrum. In contrast, our design is \emph{physics-informed}: since the eDOS can be expressed as an aggregation of pDOS, each energy bin should selectively attend to the atoms that contribute at that energy.

We introduce a structure-aware sequence decoding approach that models long-range correlations across energy bins with self attention, while grounding each bin in atom-level embeddings through source (atom-to-energy) attention.
Specifically, at each decoding step $i$, the input is a concatenation of an energy bin embedding $s_i$ and the global crystal embedding $h_x$ (a graph-level vector obtained by pooling the final-layer node embeddings $\{h_i^L\}$), forming the composite embedding $z_i = [\, s_i ~;~ h_x \,]$, where $s_i \in \mathbb{R}^{l_y}$ is a one-hot vector representing the energy index and $l_y$ is the length of the eDOS. The set $\{z_i\}_{i=1}^{l_y}$ initializes hidden states for the first self attention layer.

This mechanism operationalizes the physically motivated view of eDOS as an aggregation of pDOS with atom-projected contributions, while preserving fine-grained structure–energy dependencies. Moreover, by modeling the eDOS as a spectrum with contextual self-attention, our decoder captures smooth energy-wise correlations more effectively than static regression heads like MLPs, yielding predictions that are consistent with the underlying electronic structure.

Each decoder layer applies (i) self-attention over energy bins and (ii) source-attention to node embeddings $\{h_i^L\}$:
\begin{align}
\mathrm{Att}(Q,K,V) &= \mathrm{softmax}\!\left(\frac{QK^\top}{\sqrt{d}}\right)V, \displaybreak[0] \\
(Q,K,V) &=
\begin{cases}
(Q_{\mathrm{dec}},K_{\mathrm{dec}},V_{\mathrm{dec}}) & \text{(self)}\\
(Q_{\mathrm{dec}},K_{\mathrm{enc}},V_{\mathrm{enc}}) & \text{(source)}
\end{cases}
\end{align}
with $Q_{\mathrm{dec}}$ from decoder states and $(K_{\mathrm{enc}},V_{\mathrm{enc}})$ from $\{h_i^L\}$. The decoder outputs an unnormalized, nonnegative eDOS $\hat{y}=\{\hat{y}_i\}_{i=1}^{l_y}$ (a \texttt{Softplus} head enforces $\hat{y}_i\ge 0$). We compare it to the ground-truth $y=\{y_i\}_{i=1}^{l_y}$ using a \emph{shape-and-mass} objective. Let $\tilde{y}_i=\frac{y_i}{\sum_j y_j}$ and $\tilde{\hat{y}}_i=\frac{\hat{y}_i}{\sum_j \hat{y}_j}$ denote their sum-normalized forms.
We supervise spectral \emph{shape} with KL divergence.
\begin{equation}
\mathcal{L}_{\mathrm{Recon}}=D_{\mathrm{KL}}(\tilde{y}\,\|\,\tilde{\hat{y}})
=\sum_{i=1}^{l_y}\tilde{y}_i\log\frac{\tilde{y}_i}{\tilde{\hat{y}}_i},
\label{eq:kl_loss}
\end{equation}
The predicted total mass is read off as $\hat{Z}=\sum_i\hat{y}_i$ and supervised against the ground-truth total $Z=\sum_i y_i$. For eDOS and phDOS the total mass (number of states) has an analytic form and may instead be supplied directly; for XANES and Raman it is learned through this term. The combined objective is shown as follows:
\begin{equation}
\mathcal{L}=\mathcal{L}_{\mathrm{Recon}}+\lambda\|\hat{Z}-Z\|_2^2 .
\label{eq:total_loss}
\end{equation}
This objective treats the normalized DOS as a distribution, emphasizing correct placement of spectral mass (peaks/band edges) rather than per-bin error alone.

\vspace{-5pt}
\subsection{Constraint-Preserving Physical Reasoning}\label{sec:reason}
Given an initial prediction of the normalized eDOS from structure-aware spectrum decoding, we perform a second-stage reasoning that explicitly enforces hard physical constraints. While the previous part captures atom-projected contributions and energy-wise correlations, its output can still exhibit residual artifacts, such as blurred band edges, non-zero noise in band gap, or misplaced spectral mass. 

To correct such errors without violating physical constraints, we refine the prediction directly in probability space. We introduce a constraint-preserving physical reasoning method for eDOS, \emph{mass-conserving iterative refinement}, which treats the normalized eDOS as a probability density and updates it via a continuous transport that preserves mass by construction. This formulation enables principled redistribution of spectral mass across the energy bins, sharpening band gaps and peaks while maintaining normalization. We present the high-level formulation and algorithmic intuition here, summarize the full training and inference pipeline in \Cref{alg:deepscireasoner}, and save the detailed derivations and proofs in Appendix~\ref{app:cppr}.

Formally, we model this refinement as a transport dynamics governed by a continuity equation. Let $p_0(x) \equiv \tilde{\hat{y}}(x)$ and $p_T(x) \equiv \tilde{y}(x)$ denote the normalized eDOS corresponding to the initial prediction and the ground-truth, defined over the energy domain $x \in \mathbb{R}$ (the spectral axis $\epsilon$/$E$ of \Cref{sec:pre}, written $x$ here for the transport formulation):
$
    \int p_0(x)\,dx = \int p_T(x)\,dx = 1.
$
Our goal is to learn a transport that continuously evolves $p_0$ to $p_T$ while the total probability mass is preserved throughout the refinement trajectory. This formulation naturally admits a continuous flow interpretation, where a velocity field $f_t(x)$ governs the evolution of the normalized eDOS over time.

\vspace{-5pt}
\paragraph{Mass-Conserving Transport}
The iterative refinement must \emph{preserve total spectral mass}: for the normalized eDOS density $p(x,t)$ on an energy domain $\Omega=[a,b]$, $\int_{\Omega} p(x,t)\,dx = 1$, $\forall\, t.$
A convenient way to guarantee this is to model refinement as \emph{pure transport} of density along the energy axis, so that mass can move but cannot be created or destroyed. Let $f_\theta(x,t)$ denote a learnable velocity field. We evolve $p$ by the one-dimensional continuity equation
\begin{equation}\label{eq:continuity}
\partial_t p(x,t) + \partial_x j(x,t) \;=\; 0, \qquad j(x,t)\coloneqq p(x,t)\,f_\theta(x,t)
\end{equation}
Integrating~\Cref{eq:continuity} over $\Omega=[a,b]$ gives
\begin{equation}
\frac{d}{dt}\int_a^b p(x,t)\,dx
= -\int_a^b \partial_x j(x,t)\,dx
= -\big[j(b,t)-j(a,t)\big].
\end{equation}
Therefore, imposing \emph{no-flux} boundary conditions $j(a,t)=j(b,t)=0$ ensures $\frac{d}{dt}\int_a^b p(x,t)\,dx=0$, so that total mass is conserved throughout refinement. For $\Omega=\mathbb{R}$, it suffices to assume vanishing flux at infinity, $\lim_{x\to\pm\infty} j(x,t)=0$ (with integrable tails), so the boundary term disappears.

\vspace{-5pt}
\paragraph{Log-Density Dynamics}
Expanding~\Cref{eq:continuity} by the product rule yields
\begin{equation}
\partial_t p(x,t) + f_\theta(x,t)\,\partial_x p(x,t) + p(x,t)\,\partial_x f_\theta(x,t)=0.
\end{equation}
For points where $p(x,t)>0$, dividing by $p(x,t)$ gives the evolution of the log-density:
\begin{equation}\label{eq:log-density}
\partial_t \log p(x,t)
= -\, f_\theta(x,t)\,\partial_x \log p(x,t)\;-\;\partial_x f_\theta(x,t).
\end{equation}
Equivalently, along characteristics $x'(t)=f_\theta(x(t),t)$, the material derivative satisfies
\begin{equation}
\frac{D}{Dt}\log p(x(t),t) \;\coloneqq\; \partial_t \log p + f_\theta\,\partial_x \log p \;=\; -\,\partial_x f_\theta,
\end{equation}
showing that the log-density changes along each flow trajectory according to the local expansion/compression rate of the velocity field.

\vspace{-5pt}
\paragraph{Reference Velocity Field}
\label{sec:ref_vec}
To obtain a physically meaningful reference flow, we choose the $2$-Wasserstein displacement interpolation~\citep{mccann1997convexity} between the source and target densities $p_0$ and $p_T$ on the energy domain $[a,b]$.  In one dimension the optimal transport map $T$ that pushes $p_0$ to $p_T$ is the monotone rearrangement
\begin{equation}
T(x) = F_T^{-1}\big(F_0(x)\big),
\end{equation}
where $F_0(x)$ and $F_T(x)$ are the cumulative distribution functions (CDF) of $p_0$ and $p_T$, and $F_T^{-1}$ denotes the (generalized) inverse CDF.

We label mass elements by their initial cumulative mass $s\in[0,1]$ and define the quantile functions \( x(s)=F^{-1}(s) \).
The displacement interpolation moves each quantile along the straight segment connecting its initial and final positions,
\begin{equation}
x_t(s) = (1-t)\,x_0(s) + t\,x_T(s), \quad t\in[0,1].
\label{eq:displacement-curve}
\end{equation}
The intermediate density $p_t$ is then defined as the pushforward \( p_t = (x_t)_\sharp p_0 \), which in one dimension gives
\begin{equation}\label{eq:pt-displacement}
p_t\big(x_t(s)\big) = \frac{1}{\partial_s x_t(s)} = \Bigg[
     \frac{1-t}{p_0\big(x_0(s)\big)}
   + \frac{t}{p_T\big(x_T(s)\big)}
  \Bigg]^{-1}.
\end{equation}
Equivalently, for any $x$ and $t$ we can write $p_t(x)$ implicitly by first computing $s=F_t(x)$ with $F_t$ the CDF associated with \Cref{eq:pt-displacement}. The corresponding reference velocity field $f_t$ is defined in Lagrangian form by requiring that
\begin{equation}
\frac{d}{dt} x_t(s) = f_t\big(x_t(s)\big), \quad x_0(s)=F_0^{-1}(s).
\end{equation}
Using \Cref{eq:displacement-curve} we obtain the explicit expression
\begin{equation}
f_t\big(x_t(s)\big)
  =
\frac{d}{dt} x_t(s)
  =
x_T(s) - x_0(s),
\label{eq:reference-velocity}
\end{equation}
i.e., each mass element moves with constant velocity equal to the displacement between its initial and final OT locations. In Eulerian coordinates, for a given $(x,t)$ we recover $s=F_t(x)$ and set $f_t(x)=x_T(s)-x_0(s)$.

By construction, the pair $\{p_t,f_t\}_{t\in[0,1]}$ satisfies the continuity equation in \Cref{eq:continuity}, describing a purely advective, mass-conserving flow that transports $p_0$ to $p_T$ along the Wasserstein geodesic. We use this displacement interpolation as a physically principled reference velocity field for our subsequent constructions.

\vspace{-5pt}
\paragraph{Training}
We intend to learn a neural approximation $f_\theta(x,t)$ parameterized by $\theta$ using the analytic OT velocity field $f_t(x)$. We construct a training set of pairs $(x,t)$ sampled from the energy–time domain and supervise the network to match the reference field~\citep{hyvarinen2005estimation,lipman2022flow}:
\begin{equation}
\mathcal{L}(\theta) = \mathbb{E}_{(x,t)\sim \mathcal{D}} \big[ \| f_\theta(x,t) - f_t(x) \|^2 \big].
\label{eq:velocity-regression-loss}
\end{equation}
Since the supervision $f_t$ is from a Wasserstein-displacement interpolation that exactly satisfies the continuity equation, minimizing \Cref{eq:velocity-regression-loss} forces $f_\theta$ to approximate a physically valid, mass-conserving transport map from $p_0$ to $p_T$, without ever differentiating through the OT quantile construction or the PDE solver, enabling simulation-free training.

\vspace{-5pt}
\paragraph{Inference}
At inference time, we use the learned velocity field $f_\theta(x,t)$
to advect an initial density $p_0(x)$ forward in time by solving \Cref{eq:continuity}
over $t\in[0, T]$.  If $f_\theta$ coincided exactly with the reference
field $f_t$, the solution $p_\theta(\cdot,t)$ would follow the same
reference Wasserstein geodesic.

For numerical stability, we implement this evolution in the Eulerian domain via $F_\theta(x,t) = \int_{-\infty}^x p_\theta(u,t)\,du,$ which obeys the equivalent first-order PDE
\begin{equation}
\label{eq:cdf-pde}
\begin{gathered}
    \partial_t F_\theta(x,t) + f_\theta(x,t)\, \partial_x F_\theta(x,t) = 0, \quad F_\theta(x,0) = F_0(x).
\end{gathered}
\end{equation}
Integrating \Cref{eq:cdf-pde} from $t=0$ to $T$ with the initial
CDF $F_0(x)$ yields $F_\theta(x, T)$, and the refined target
density $\hat{p}_T(x)$ can be recovered as follows:
\begin{equation}
\hat{p}_T(x)
\;:=\;
p_\theta(x, T)
=
\partial_x F_\theta(x, T).
\label{eq:refined-target-density}
\end{equation}
This Eulerian formulation provides a stable and deterministic
procedure for transporting $p_0$ to a refined density $\hat{p}_T$
while strictly conserving the total probability mass and remaining
consistent with the learned OT-informed dynamics $f_\theta$.

\vspace{-5pt}
\paragraph{Discretization}
We discretize the energy domain $\Omega=[a,b]$ on a uniform grid $x_i=a+i\Delta x$ ($i=0,\dots,N_x-1$, where $N_x=l_y$ is the number of spectral bins) with $\Delta x=(b-a)/(N_x-1)$, and the refinement horizon $[0,T]$ with time step $\Delta t=T/(N_t-1)$. Let $F_i^k \approx F_\theta(x_i,t_k)$. We evolve~\Cref{eq:cdf-pde} using forward Euler in time and a first-order upwind derivative in space:
\begin{equation}
F_i^{k+1}=F_i^k-\Delta t\, f_\theta(x_i,t_k)\,D_x^{\mathrm{up}}F_i^k.
\label{eq:discrete-cdf-update}
\end{equation}
This monotone upwind scheme preserves CDF ordering and (up to rounding) conserves total mass. After $N_t-1$ update steps, we recover the refined density at $t=T$ by finite differences,
\begin{equation}
\hat{p}_T(x_i)=\frac{F_i^{N_t-1}-F_{i-1}^{N_t-1}}{\Delta x},\quad i=1,\ldots,N_x-1,
\end{equation}
with a final renormalization so that $\sum_i \hat{p}_T(x_i)\Delta x=1$.

\vspace{-5pt}
\paragraph{Amortized Inference}
The preceding formulation describes how we learn a single velocity field that transports one specific source–target pair of densities $(p_0, p_T)$.
In practice, the training set contains $N$ pairs of normalized eDOS ``distributions" $\{(p_0^{(i)}, p_T^{(i)})\}_{i=1}^N$, each corresponding to a different crystal structure.
To enable generalization across crystals rather than fitting a separate field for each, we adopt an \emph{amortized inference} strategy.
Specifically, a shared neural network parameterized by $\theta$ is trained to predict the velocity field $f_\theta(x, t; \mathcal{C})$, conditioned on contextual features $\mathcal{C}$ (e.g., crystal descriptors or embeddings) that identify each sample.
This amortized inference enables a unified transport operator capable of learning continuous density trajectories for any crystal-specific eDOS pair.

\section{Experiments}\label{sec:exp}
In this section, we address four research questions:
(\textbf{RQ1}) Does DeepSciReasoner improve the prediction performance of structured scientific spectra (eDOS, phDOS, XANES, Raman) in general?
(\textbf{RQ2}) Does DeepSciReasoner sharpen the spectral edges and improve identification of spectral gaps?
(\textbf{RQ3}) Does DeepSciReasoner exhibit improved data-scaling behavior, yielding stronger sample efficiency?
(\textbf{RQ4}) Does the proposed constrained physical reasoning contribute measurable gains in spectrum prediction performance?

\vspace{-5pt}
\subsection{Experimental Setup}
Here we describe the necessary experimental setup. For additional details, please refer to \Cref{app:hyper}, and we report computational cost in \Cref{app:comp_cost}.

\begin{table*}[ht!] 
\caption{Performance of eDOS prediction on two benchmark datasets. We report mean ± one standard deviation. Darkly shaded cells indicate the best results and lightly shaded cells indicate the second best.}
\label{tab:edos}
\begin{center}
\begin{sc}
\vspace{-5pt}
\resizebox{\textwidth}{!}{%
\begin{tabular}{l*{8}{c}}
\toprule 
& \multicolumn{4}{c}{\textbf{CalTech}} & \multicolumn{4}{c}{\textbf{TRI}} \\ 
\cmidrule(lr){2-5} \cmidrule(lr){6-9}
Method 
& R$^2 \uparrow$ 
& MAE $\downarrow$ 
& MSE $\downarrow$ 
& WD $\downarrow$ 
& R$^2 \uparrow$ 
& MAE $\downarrow$ 
& MSE $\downarrow$ 
& WD $\downarrow$ \\
\midrule 

E3NN       
& 0.748\scriptsize{$\pm$0.002} & 2.421\scriptsize{$\pm$0.003} & 27.351\scriptsize{$\pm$0.629} & 0.261\scriptsize{$\pm$0.005} 
& 0.669\scriptsize{$\pm$0.004} & 2.945\scriptsize{$\pm$0.005} & 36.535\scriptsize{$\pm$0.532} & 0.445\scriptsize{$\pm$0.004} \\

GATGNN         
& \cellcolor{second}0.772\scriptsize{$\pm$0.001} & 2.272\scriptsize{$\pm$0.016} & 24.831\scriptsize{$\pm$0.585} & 0.230\scriptsize{$\pm$0.004} 
& 0.702\scriptsize{$\pm$0.004} & 2.724\scriptsize{$\pm$0.004} & 33.811\scriptsize{$\pm$0.420} & \cellcolor{second}0.370\scriptsize{$\pm$0.003} \\

CrystalFormer  
& 0.765\scriptsize{$\pm$0.006} 
& 2.344\scriptsize{$\pm$0.016} 
& 26.138\scriptsize{$\pm$0.477} 
& 0.248\scriptsize{$\pm$0.012} 
& 0.695\scriptsize{$\pm$0.010} 
& 2.869\scriptsize{$\pm$0.082} 
& 34.584\scriptsize{$\pm$0.581} 
& 0.399\scriptsize{$\pm$0.017} \\

\midrule

Mat2Spec       
& 0.743\scriptsize{$\pm$0.002} & 2.226\scriptsize{$\pm$0.010} & 27.955\scriptsize{$\pm$0.137} & 0.228\scriptsize{$\pm$0.013} 
& 0.677\scriptsize{$\pm$0.002} & 2.719\scriptsize{$\pm$0.072} & 36.628\scriptsize{$\pm$0.105} & 0.395\scriptsize{$\pm$0.009} \\

DOSTransformer 
& 0.771\scriptsize{$\pm$0.002} & \cellcolor{second}2.023\scriptsize{$\pm$0.032} & \cellcolor{second}24.675\scriptsize{$\pm$0.354} & \cellcolor{second}0.212\scriptsize{$\pm$0.040} 
& \cellcolor{second}0.716\scriptsize{$\pm$0.003} & \cellcolor{second}2.455\scriptsize{$\pm$0.080} & \cellcolor{second}32.208\scriptsize{$\pm$0.288} & 0.375\scriptsize{$\pm$0.014} \\

\midrule

DeepSciReasoner      
& \cellcolor{best}\textbf{0.815\scriptsize{$\pm$0.007}} & \cellcolor{best}\textbf{1.983\scriptsize{$\pm$0.021}} & \cellcolor{best}\textbf{19.612\scriptsize{$\pm$0.225}} & \cellcolor{best}\textbf{0.191\scriptsize{$\pm$0.017}} 
& \cellcolor{best}\textbf{0.761\scriptsize{$\pm$0.009}} & \cellcolor{best}\textbf{2.357\scriptsize{$\pm$0.046}} & \cellcolor{best}\textbf{27.162\scriptsize{$\pm$0.563}} & \cellcolor{best}\textbf{0.305\scriptsize{$\pm$0.011}} \\

\bottomrule 
\end{tabular}}
\end{sc}
\end{center}
\vspace{-5pt}
\end{table*}

\begin{table*}[t!]
\caption{Generalization to other structured spectra. Performance on XANES-full (full multi-absorber), XANES-O (oxygen K-edge subset), Raman, and phDOS benchmarks. Darkly shaded cells denote the best results and lightly shaded cells denote the second best. \textbf{Abbreviations:} DSR = DeepSciReasoner.}
\label{tab:gen}
\begin{center}
\begin{sc}
\vspace{-5pt}
\resizebox{\textwidth}{!}{%
\begin{tabular}{l*{10}{c}}
\toprule
& \multicolumn{2}{c}{\textbf{XANES-full}}
& \multicolumn{2}{c}{\textbf{XANES-O}}
& \multicolumn{2}{c}{\textbf{Raman}}
& \multicolumn{4}{c}{\textbf{phDOS}} \\
\cmidrule(lr){2-3} \cmidrule(lr){4-5} \cmidrule(lr){6-7} \cmidrule(lr){8-11}
Method
& R$^2 \uparrow$ & WD $\downarrow$
& R$^2 \uparrow$ & WD $\downarrow$
& R$^2 \uparrow$ & WD $\downarrow$
& R$^2 \uparrow$ & MAE $\downarrow$ & MSE $\downarrow$ & WD $\downarrow$ \\
\midrule
E3NN
& 0.650\scriptsize{$\pm$0.012} & 0.108\scriptsize{$\pm$0.003}
& 0.381\scriptsize{$\pm$0.011} & 0.099\scriptsize{$\pm$0.004}
& 0.096\scriptsize{$\pm$0.005} & 0.451\scriptsize{$\pm$0.019}
& 0.506\scriptsize{$\pm$0.015} & 0.082\scriptsize{$\pm$0.003} & 0.026\scriptsize{$\pm$0.001} & 0.103\scriptsize{$\pm$0.004} \\
GATGNN
& 0.648\scriptsize{$\pm$0.005} & \cellcolor{second}0.104\scriptsize{$\pm$0.001}
& 0.293\scriptsize{$\pm$0.013} & 0.108\scriptsize{$\pm$0.002}
& 0.115\scriptsize{$\pm$0.005} & 0.526\scriptsize{$\pm$0.008}
& 0.270\scriptsize{$\pm$0.025} & 0.106\scriptsize{$\pm$0.002} & 0.038\scriptsize{$\pm$0.003} & 0.155\scriptsize{$\pm$0.008} \\
MEGNet
& \cellcolor{second}0.657\scriptsize{$\pm$0.008} & 0.107\scriptsize{$\pm$0.002}
& 0.347\scriptsize{$\pm$0.020} & 0.102\scriptsize{$\pm$0.006}
& 0.045\scriptsize{$\pm$0.004} & 0.449\scriptsize{$\pm$0.024}
& 0.538\scriptsize{$\pm$0.014} & 0.080\scriptsize{$\pm$0.003} & 0.024\scriptsize{$\pm$0.001} & 0.094\scriptsize{$\pm$0.004} \\
M3GNet
& 0.648\scriptsize{$\pm$0.010} & 0.107\scriptsize{$\pm$0.002}
& \cellcolor{second}0.386\scriptsize{$\pm$0.012} & \cellcolor{second}0.096\scriptsize{$\pm$0.005}
& 0.083\scriptsize{$\pm$0.003} & \cellcolor{second}0.426\scriptsize{$\pm$0.021}
& \cellcolor{second}0.543\scriptsize{$\pm$0.012} & \cellcolor{second}0.075\scriptsize{$\pm$0.003} & \cellcolor{second}0.023\scriptsize{$\pm$0.001} & \cellcolor{second}0.075\scriptsize{$\pm$0.003} \\

\midrule
DSR
& \cellcolor{best}\textbf{0.764}\scriptsize{$\pm$0.009} & \cellcolor{best}\textbf{0.098}\scriptsize{$\pm$0.003}
& \cellcolor{best}\textbf{0.474}\scriptsize{$\pm$0.010} & \cellcolor{best}\textbf{0.093}\scriptsize{$\pm$0.003}
& \cellcolor{best}\textbf{0.264}\scriptsize{$\pm$0.002} & \cellcolor{best}\textbf{0.423}\scriptsize{$\pm$0.017}
& \cellcolor{best}\textbf{0.636\scriptsize{$\pm$0.013}} & \cellcolor{best}\textbf{0.062\scriptsize{$\pm$0.004}} & \cellcolor{best}\textbf{0.021\scriptsize{$\pm$0.002}} & \cellcolor{best}\textbf{0.061\scriptsize{$\pm$0.002}} \\
\bottomrule
\end{tabular}}
\end{sc}
\end{center}
\vspace{-10pt}
\end{table*}

\vspace{-5pt}
\paragraph{Dataset Curation}
For \textbf{eDOS}, we evaluate on two benchmark datasets, CalTech~\citep{Kong2022} and TRI~\citep{mueller2024electrochemical}, both derived from the Materials Project~\citep{Jain2013}. Unlike prior work that removes the energy interval between the valence-band maximum (VBM) and conduction-band minimum (CBM) for gapped materials, we retain the full eDOS so that band gaps are represented explicitly. CalTech applies a binning procedure to the raw DOS, where the points within each bin are averaged and assigned to the energy of the bin center, yielding 128 energy bins over $[-4,4]$~eV. TRI instead applies Gaussian smoothing ($\sigma=2$~eV) and discretizes onto 256 bins over $[-5,5]$~eV. This full-spectrum setting preserves the density behavior near band edges, which is more challenging.

For \textbf{phDOS}, we follow~\citet{chen2021direct} and use their curated DFPT phDOS dataset of 1{,}521 crystalline materials drawn from Materials Project; each spectrum is discretized on a 51-bin grid spanning $[0,1000]$~cm$^{-1}$ with sum-normalization. All datasets use an 80/10/10 train/val/test split. Full step-by-step processing and gating rules are deferred to \Cref{app:data}.

For \textbf{XANES}, we curate K-edge spectra from the Materials Project XAS database~\citep{Jain2013,mathew2018xas}, retaining 49,895 (mpid, absorber) pairs after gating, and discretize each spectrum on a 128-bin grid over $[-5,+40]$~eV (relative to the per-element K-edge $E_0$) with Gaussian smearing ($\sigma=0.5$~eV) and sum-normalization. Because the same crystal graph can correspond to spectra from different absorbing elements, we additionally condition the decoder on the absorber atomic number. We also construct a single-absorber oxygen subset (5{,}094 mpids) so that baselines built around the one-spectrum-per-material convention can be compared on equal footing. 

For \textbf{Raman}, we use the high-throughput DFPT database of~\citet{bagheri2023raman}: powder Raman intensities are computed per mode via the Placzek formula~\citep{placzek1934rayleigh}, placed as Gaussians ($\sigma=10$~cm$^{-1}$) on a 128-bin grid over $[0,1500]$~cm$^{-1}$, and sum-normalized, yielding 2{,}283 mpids after gating centrosymmetric crystals (which have no Raman-active modes).

\vspace{-5pt}
\paragraph{Baselines}
We benchmark against baselines that fall into two main families. E3NN~\citep{geiger2022e3nn}, GATGNN~\citep{louis2020graph}, MEGNet~\citep{Chen2019}, M3GNet~\citep{chen2022m3gnet}, and CrystalFormer~\citep{wang2024conformal} are all structure encoders that learn a representation of a crystal graph and then map it to targets; they mainly differ in inductive bias. In contrast, Mat2Spec~\citep{Kong2022} and DOSTransformer~\citep{lee2023density} are crystal-to-sequence models that treat the spectrum as a sequence across the spectral bins. E3NN and GATGNN are evaluated on all targets; CrystalFormer, Mat2Spec, and DOSTransformer are eDOS-specific (\Cref{tab:edos}), whereas MEGNet and M3GNet serve as encoders for the generalization benchmarks (\Cref{tab:gen}). More details in \Cref{app:base}.

\vspace{-5pt}
\paragraph{Evaluation Metrics}
We evaluate performance with task-specific metrics: (i) for overall eDOS and phDOS prediction, we report mean absolute error (MAE), mean squared error (MSE), coefficient of determination (R$^2$), and Wasserstein distance (WD), all computed against the DFT ground-truth eDOS; (ii) for VB gap identification, we report precision, recall, F1 score, and Matthews correlation coefficient (MCC); and (iii) for overall XANES and Raman prediction, we report coefficient of determination (R$^2$) and Wasserstein distance (WD). More details in \Cref{app:eval}.

\vspace{-5pt}
\subsection{Results and Analysis} \label{sec:res}
\vspace{-5pt}
\paragraph{(RQ1) Overall Prediction of Structured Spectra}
Table \ref{tab:edos} summarizes full-spectrum eDOS prediction accuracy on CalTech and TRI. Across all metrics, DeepSciReasoner consistently achieves the best performance on both datasets. Notably, on CalTech, it reduces MSE by 20.52\% relative to the strongest baseline, highlighting the benefit of physical reasoning beyond neural sequence decoding alone. In contrast, neural encoders such as E3NN, GATGNN, and CrystalFormer show limited ability to capture fine-grained spectral geometry, while crystal-to-sequence methods improve MAE but still struggle with global distributional alignment. These results demonstrate that combining neural spectrum decoder with constraint-preserving physical reasoning yields more accurate and physically faithful predictions. For details of the qualitative comparison (visualization), please refer to \Cref{app:viz}.

\Cref{tab:gen} reports performance on four additional structured-spectrum benchmarks spanning XANES, Raman, and phDOS. DeepSciReasoner achieves the best result across all XANES and Raman metrics, improving XANES-full $R^2$ from $0.657$ for the strongest baseline to $0.764$, and reducing WD from $0.104$–$0.108$ to $0.098$. On the oxygen K-edge XANES subset, it similarly leads both metrics, indicating that the gains persist under a more chemically restricted evaluation setting. On Raman spectra, which are sparse and sharply peaked, DeepSciReasoner improves $R^2$ substantially from the strongest baseline value of $0.115$ to $0.264$ while also achieving the lowest WD, showing that the framework transfers beyond density-of-states-like targets to vibrational response spectra. On phDOS, DeepSciReasoner achieves the best performance across all four metrics, outperforming the strongest baseline (M3GNet) on $R^2$, MAE, MSE, and WD. Overall, these results support the central thesis that structure-aware spectrum decoding coupled with mass-conserving refinement provides a reusable blueprint for structured scientific spectra, transferring without architectural surgery from eDOS to electronic-excitation, vibrational, and lattice-dynamical spectra.

\vspace{-5pt}
\paragraph{(RQ2) Sharp Gap Identification}
Here, we use valence-band (VB) gap as the example to evaluate how well the model identifies sharp edges and gaps, because inferring VB gap is a practical screening task. It directly tests whether a predicted eDOS is reliable enough to recover rare, near–Fermi-level suppressed-state regions that are easy to wash out under smoothing or regression noise. In applications, this enables high-throughput triage of candidate materials for follow-up experiments, since VB gaps are linked to transport-relevant behavior while being too rare to find efficiently by brute-force calculations. 

\Cref{tab:vb_summary} reports VB-gap identification performance, a highly imbalanced setting where only a small fraction of materials exhibit near–Fermi-level VB gaps. DeepSciReasoner consistently achieves the strongest performance across all applicable metrics on both datasets with substantial gains. Neural encoders (E3NN, GATGNN, CrystalFormer) struggle due to their indirect mapping from global representations, while crystal-to-sequence models improve precision but remain recall-limited. In contrast, combined deep learning and constrained reasoning, DeepSciReasoner yields eDOS predictions that are not only globally accurate but also sufficiently sharp and well-calibrated to support downstream identification of rare, functionally relevant electronic features.

\begin{table*}[htbp] 
\caption{VB-gap identification performance. We report mean ± one standard deviation. Darkly shaded cells denote the best results and lightly shaded cells denote the second best among applicable methods.}
\label{tab:vb_summary}
\begin{center} 
\begin{sc}
\vspace{-5pt}
\resizebox{\textwidth}{!}{%
\begin{tabular}{l*{8}{c}}
\toprule
& \multicolumn{4}{c}{\textbf{CalTech}} 
& \multicolumn{4}{c}{\textbf{TRI}} \\
\cmidrule(lr){2-5} \cmidrule(lr){6-9}
Method 
& Prec. $\uparrow$ & Recall $\uparrow$ & F1 $\uparrow$ & MCC $\uparrow$ 
& Prec. $\uparrow$ & Recall $\uparrow$ & F1 $\uparrow$ & MCC $\uparrow$ \\
\midrule

E3NN       
& 0.672{\scriptsize$\pm$0.035} & 0.366{\scriptsize$\pm$0.013} & 0.474{\scriptsize$\pm$0.003} & 0.479{\scriptsize$\pm$0.011} 
& 0.647{\scriptsize$\pm$0.052} & 0.314{\scriptsize$\pm$0.071} & 0.423{\scriptsize$\pm$0.028} & 0.442{\scriptsize$\pm$0.012} \\

GATGNN         
& 0.699{\scriptsize$\pm$0.033} & 0.430{\scriptsize$\pm$0.008} & 0.532{\scriptsize$\pm$0.003} & 0.532{\scriptsize$\pm$0.009}
& \cellcolor{second}0.671{\scriptsize$\pm$0.049} & 0.362{\scriptsize$\pm$0.053} & 0.476{\scriptsize$\pm$0.021} & 0.489{\scriptsize$\pm$0.008} \\

CrystalFormer  
& 0.654{\scriptsize$\pm$0.023} 
& 0.443{\scriptsize$\pm$0.031} 
& 0.532{\scriptsize$\pm$0.022} 
& 0.542{\scriptsize$\pm$0.013} 
& 0.616{\scriptsize$\pm$0.037} 
& 0.371{\scriptsize$\pm$0.009} 
& 0.465{\scriptsize$\pm$0.024} 
& 0.472{\scriptsize$\pm$0.031} \\
\midrule

Mat2Spec       
& \cellcolor{second}0.726{\scriptsize$\pm$0.018} & \cellcolor{second}0.551{\scriptsize$\pm$0.007} & \cellcolor{second}0.629{\scriptsize$\pm$0.004} & \cellcolor{second}0.611{\scriptsize$\pm$0.019}
& 0.601{\scriptsize$\pm$0.024} & \cellcolor{second}0.429{\scriptsize$\pm$0.016} & \cellcolor{second}0.509{\scriptsize$\pm$0.017} & \cellcolor{second}0.508{\scriptsize$\pm$0.014} \\

DOSTransformer 
& 0.684{\scriptsize$\pm$0.027} & 0.532{\scriptsize$\pm$0.009} & 0.594{\scriptsize$\pm$0.014} & 0.584{\scriptsize$\pm$0.051}
& 0.573{\scriptsize$\pm$0.073} & 0.329{\scriptsize$\pm$0.020} & 0.416{\scriptsize$\pm$0.052} & 0.422{\scriptsize$\pm$0.066} \\

\midrule

DeepSciReasoner      
& \cellcolor{best}\textbf{0.753{\scriptsize$\pm$0.021}} 
& \cellcolor{best}\textbf{0.637{\scriptsize$\pm$0.014}} 
& \cellcolor{best}\textbf{0.704{\scriptsize$\pm$0.011}} 
& \cellcolor{best}\textbf{0.679{\scriptsize$\pm$0.022}} 
& \cellcolor{best}\textbf{0.803{\scriptsize$\pm$0.035}} 
& \cellcolor{best}\textbf{0.571{\scriptsize$\pm$0.064}} 
& \cellcolor{best}\textbf{0.648{\scriptsize$\pm$0.047}} 
& \cellcolor{best}\textbf{0.645{\scriptsize$\pm$0.063}} \\

\bottomrule
\end{tabular}}
\end{sc}
\end{center}
\vspace{-12pt}
\end{table*}

\begin{table*}[htbp] 
\caption{Ablation study of DeepSciReasoner on eDOS prediction. We report mean $\pm$ one standard deviation. Darkly shaded cells denote the best results and lightly shaded cells denote the second best. \textbf{Abbreviations:} DSR = DeepSciReasoner; NR = w/o constraint-preserving physical reasoning; NDNR = w/o both components.}
\label{tab:abl}
\begin{center}
\begin{sc}
\vspace{-5pt}
\resizebox{\textwidth}{!}{%
\begin{tabular}{l*{8}{c}}
\toprule 
& \multicolumn{4}{c}{\textbf{CalTech}} & \multicolumn{4}{c}{\textbf{TRI}} \\ 
\cmidrule(lr){2-5} \cmidrule(lr){6-9}
Method 
& R$^2 \uparrow$ & MAE $\downarrow$ & MSE $\downarrow$ & WD $\downarrow$ 
& R$^2 \uparrow$ & MAE $\downarrow$ & MSE $\downarrow$ & WD $\downarrow$ \\
\midrule 

DSR (full)
& \cellcolor{best}\textbf{0.815{\scriptsize$\pm$0.007}} & \cellcolor{best}\textbf{1.983{\scriptsize$\pm$0.021}} & \cellcolor{best}\textbf{19.612{\scriptsize$\pm$0.225}} & \cellcolor{best}\textbf{0.191{\scriptsize$\pm$0.017}} 
& \cellcolor{best}\textbf{0.761{\scriptsize$\pm$0.009}} & \cellcolor{best}\textbf{2.357{\scriptsize$\pm$0.046}} & \cellcolor{best}\textbf{27.162{\scriptsize$\pm$0.563}} & \cellcolor{best}\textbf{0.305{\scriptsize$\pm$0.011}} \\

DSR-NR
& \cellcolor{second}0.807{\scriptsize$\pm$0.003} & \cellcolor{second}2.061{\scriptsize$\pm$0.020} & \cellcolor{second}20.990{\scriptsize$\pm$0.373} & \cellcolor{second}0.207{\scriptsize$\pm$0.005}
& \cellcolor{second}0.742{\scriptsize$\pm$0.005} & \cellcolor{second}2.420{\scriptsize$\pm$0.018} & \cellcolor{second}29.387{\scriptsize$\pm$0.641} & \cellcolor{second}0.329{\scriptsize$\pm$0.004} \\

DSR-NDNR
& 0.779{\scriptsize$\pm$0.001} & 2.259{\scriptsize$\pm$0.003} & 24.726{\scriptsize$\pm$0.174} & 0.223{\scriptsize$\pm$0.007}
& 0.708{\scriptsize$\pm$0.003} & 2.701{\scriptsize$\pm$0.005} & 33.196{\scriptsize$\pm$0.136} & 0.372{\scriptsize$\pm$0.003} \\

\bottomrule 
\end{tabular}}
\end{sc}
\end{center}
\vspace{-18pt}
\end{table*}

\begin{figure}[ht]
    \centering
    \includegraphics[width=0.95\linewidth]{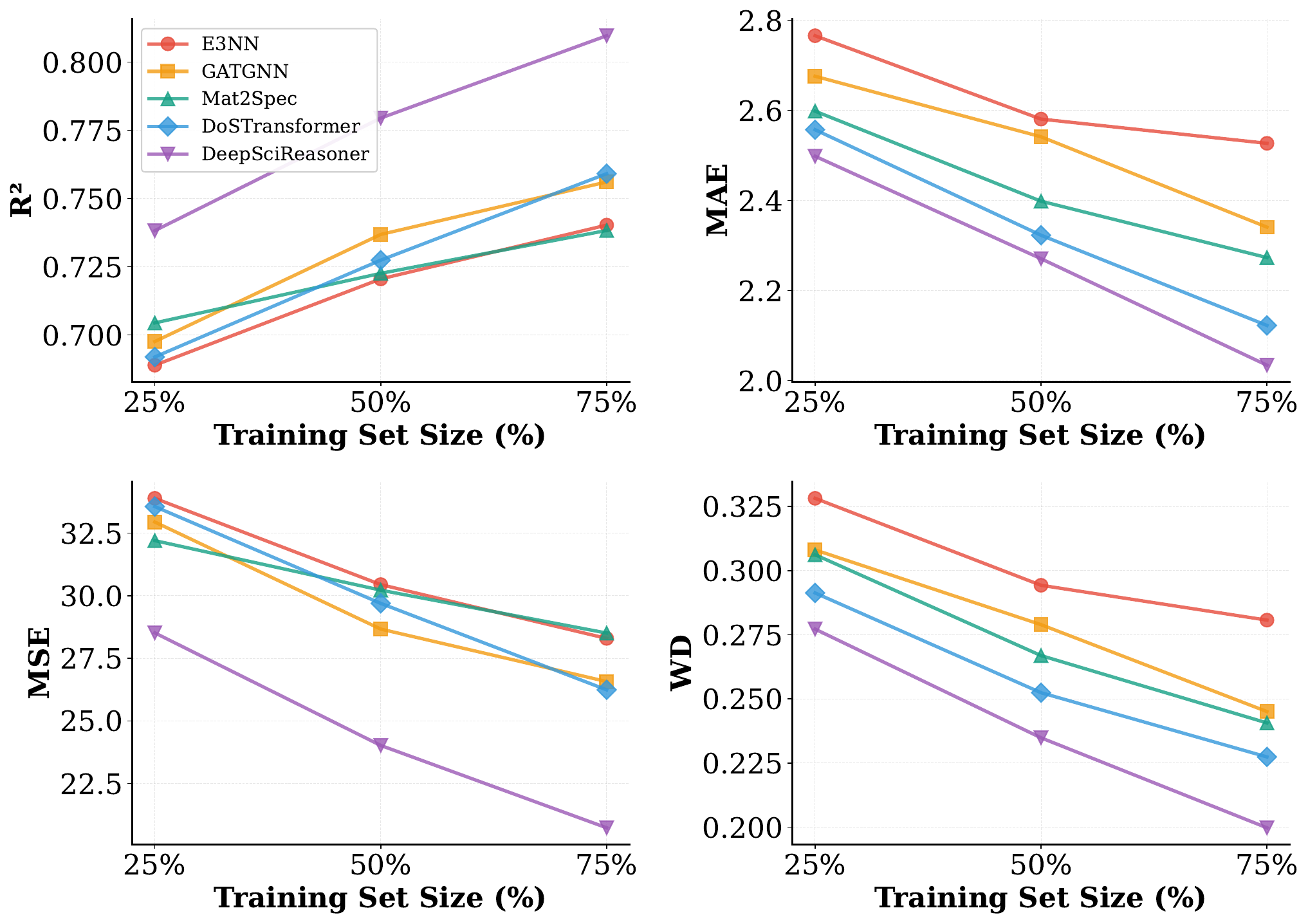}
    \caption{Data scaling behavior on the CalTech dataset. Performance of different methods as a function of training set size (25\%, 50\%, 75\%) on R$^2$, MAE, MSE, and WD.} \label{fig:sca}
    \vspace{-12pt}
\end{figure}

\vspace{-5pt}
\paragraph{(RQ3) Data Scaling Behavior}
\Cref{fig:sca} shows the scaling behavior of all methods on the CalTech dataset as the training set size increases from 25\% to 75\%. Across all metrics, DeepSciReasoner consistently outperforms baselines at every data regime and exhibits a steeper improvement trend with additional data. In particular, DeepSciReasoner achieves substantially better performance in the low-data (25\%) regime, indicating stronger sample efficiency due to its reasoning capability. As training data increases, the performance gap widens rather than narrows, suggesting that DeepSciReasoner enables more effective utilization of additional supervision. In contrast, baselines improve more slowly and remain consistently worse. The results on the TRI dataset are in \Cref{app:sca_tri}.

\vspace{-5pt}
\paragraph{(RQ4) Ablation Study}
\Cref{tab:abl} quantifies the contribution of DeepSciReasoner’s two core components: structure-aware spectrum decoding and constraint-preserving physical reasoning. Removing the physical reasoning module alone (DSR-NR) consistently degrades performance across all metrics on both eDOS benchmarks, indicating that mass-conserving refinement provides stable gains in accuracy and spectral alignment. Removing both components (DSR-NDNR) leads to substantially larger drops, confirming that structure-aware decoding is essential for capturing the overall spectral profile. The degradation pattern also clarifies their complementary roles: constraint-preserving reasoning yields moderate but consistent improvements by refining spectral geometry and enforcing conservation, whereas structure-aware decoding contributes more strongly to global predictive accuracy. For example, relative to DSR-NR, removing structure-aware decoding further increases MSE by 17.80\% on CalTech and WD by 13.07\% on TRI, highlighting its importance for accurate spectrum reconstruction.

\vspace{-5pt}
\section{Conclusion}
We introduce DeepSciReasoner, a general framework for deep scientific reasoning under hard physical constraints, instantiated here for structure-aware spectrum prediction. Using eDOS as the prototypical setting, DeepSciReasoner couples atom-to-energy spectrum decoding with mass-conserving transport-based refinement to produce predictions that are both quantitatively accurate and physically faithful: they preserve total-state conservation while sharpening band edges and other spectral features. Extensive experiments demonstrate state-of-the-art performance across full-spectrum prediction tasks spanning eDOS, phDOS, XANES, and Raman spectra, as well as robust valence-band gap identification and improved data efficiency relative to prior methods. More broadly, this work highlights the value of integrating expressive neural models with principled physical reasoning, offering a reusable blueprint for learning structured scientific spectra governed by conservation laws and geometric constraints.

\vspace{-5pt}
\section{Limitation}
Despite its broad applicability, DeepSciReasoner has several limitations. First, the framework relies on high-quality simulated spectra for supervision, and its performance may inherit biases from the underlying first-principles calculations, preprocessing choices, and spectral discretization. Second, while the mass-conserving refinement enforces normalization and improves spectral geometry, not all spectrum types have equally well-defined conservation laws or analytic total intensities, so additional domain-specific constraints may be needed for broader deployment. Third, our evaluation focuses on benchmark datasets derived primarily from crystalline inorganic materials; extending the framework to experimental spectra, molecular systems, defective structures, and out-of-distribution chemistries remains an important direction for future work.

\section*{Acknowledgements}
This project was partially supported by an AI2050 Senior Fellowship, a Schmidt Sciences program; the National Science Foundation (NSF); the National Institute of Food and Agriculture (USDA/NIFA; 2023-67021-39829); and the Air Force Office of Scientific Research (AFOSR; FA9550-23-1-0322).

\section*{Impact Statement}
This work advances AI for scientific discovery by introducing a general framework for reasoning over structured scientific spectra under hard physical constraints. By improving the accuracy and physical fidelity of electronic density-of-states prediction, the proposed approach has the potential to accelerate high-throughput materials screening and reduce reliance on expensive first-principles simulations, supporting applications in energy, electronics, and materials design. The methods developed here operate entirely on existing simulation data and do not introduce new mechanisms for data collection or deployment that raise privacy, safety, or fairness concerns. More broadly, this work promotes the responsible integration of physical laws into learning systems, encouraging models whose predictions are both interpretable and physically consistent, which we view as a positive direction for the reliable use of AI in scientific domains.

\bibliography{icml2026}
\bibliographystyle{icml2026}

\newpage
\appendix
\onecolumn

\section{Physical Background of Structured Scientific Spectra}
\label{app:edos}

This appendix provides physical background on the four structured scientific spectra studied in this work: the electronic density of states (eDOS), the phonon density of states (phDOS), X-ray absorption near-edge structure (XANES), and Raman spectra. Despite originating from distinct physical processes, these spectra share a common mathematical structure: each is a nonnegative, energy- or frequency-resolved spectrum that obeys hard physical constraints (nonnegativity, sum-normalization, and conservation laws derived from spectral sum rules) and exhibits geometric features (sharp band edges, gaps, Van Hove singularities, edge jumps, and sparse vibrational lines) whose accurate reproduction is required for the predictions to carry physical meaning. We summarize the definitions, physical content, conservation laws, and characteristic features of each spectrum in turn, motivating the unified treatment adopted by DeepSciReasoner.

\subsection{Electronic Density of States (eDOS)}
\label{app:edos_sub}

The \emph{electronic density of states} (eDOS) is a fundamental quantity in solid-state physics that characterizes how electronic states are distributed in energy within a crystalline material~\citep{kittel2018introduction}. For a periodic system with single-particle energy eigenvalues $\{\varepsilon_n(\mathbf{k})\}$ indexed by band $n$ and crystal momentum $\mathbf{k}$, the eDOS is defined as
\begin{equation}
D(E)
\;=\;
\frac{1}{V_{\mathrm{BZ}}}
\sum_{n}
\int_{\mathrm{BZ}}
\delta\!\left(E - \varepsilon_n(\mathbf{k})\right)
\, d\mathbf{k},
\label{eq:edos_def}
\end{equation}
where the integral is taken over the Brillouin zone (BZ) and $V_{\mathrm{BZ}}$ denotes its volume. Intuitively, $D(E)\,dE$ measures the number of available electronic states whose energies lie within the interval $[E, E+dE]$.

The eDOS provides a compact yet information-rich summary of a material's electronic structure. Many physical properties, including electrical conductivity, optical absorption, heat capacity, and magnetic response, depend on integrals or moments of the eDOS weighted by appropriate occupation statistics~\citep{hook2013solid}. In particular, the behavior of the eDOS near the Fermi level governs low-temperature transport, while its global structure reflects orbital hybridization and bonding characteristics encoded by the crystal geometry.

A defining feature of the eDOS is the presence of \emph{band edges} and \emph{band gaps}. For semiconductors and insulators, the eDOS vanishes over a finite energy interval separating the valence and conduction bands. The detailed shape of the eDOS near the valence-band maximum (VBM) and conduction-band minimum (CBM) determines effective carrier masses, carrier densities, and optical transition strengths~\citep{lee2016computational}. Accurately capturing these band edges, rather than merely predicting a scalar band gap, is therefore essential for faithful electronic structure modeling.

In addition to the fundamental band gap between the valence and conduction bands, the electronic density of states may exhibit valence-band gaps (\emph{VB gaps}), defined as energy intervals with vanishing density of states within the valence-band manifold below the Fermi level~\citep{harrison1989electronic}. Such VB gaps arise from orbital, bonding--antibonding, crystal-field, or correlation-induced splittings, and reflect intrinsic electronic structure beyond the global band gap~\citep{cox1987electronic}. Unlike the VBM--CBM band gap, VB gaps do not separate occupied from unoccupied states, but instead encode internal organization of occupied electronic states and are therefore important for faithfully representing the fine structure of the valence-band eDOS.

Beyond band gaps and VB gaps, the eDOS often exhibits sharp peaks known as \emph{van Hove singularities}~\citep{van1953occurrence}. These arise from critical points in the band structure where the group velocity vanishes, i.e.,
$\nabla_{\mathbf{k}} \varepsilon_n(\mathbf{k}) = \mathbf{0}$.
Van Hove singularities encode the topology of the electronic bands and are associated with enhanced electronic responses, influencing phenomena such as superconductivity, magnetism, and electronic instabilities. Resolving the location and amplitude of these peaks is particularly challenging, as they are highly sensitive to long-range crystal symmetry and dispersion relations.

In practice, the eDOS is computed using density functional theory (DFT) by discretizing the Brillouin zone and approximating the Dirac delta function with a smoothing kernel, such as a Gaussian or Savitzky--Golay filter. The resulting eDOS is represented on a finite energy grid and often normalized so that
\begin{equation}
\int_{a}^{b} D(E)\, dE
\;=\;
N_{\mathrm{states}},
\label{eq:edos_mass}
\end{equation}
where $[a,b]$ denotes the energy window of interest and $N_{\mathrm{states}}$ is the total number of electronic states within that window. Conservation of total spectral mass is a fundamental physical constraint that any predictive model of the eDOS must respect.

Importantly, the total eDOS admits a decomposition into \emph{atom-projected} or \emph{orbital-projected} densities of states (pDOS), which quantify the contribution of individual atomic species or orbitals to the electronic states at each energy. Such decompositions provide interpretability and link spectral features to local chemical environments, allowing prominent peaks in the eDOS to be traced back to specific atoms or bonding motifs.

Taken together, the eDOS constitutes a \emph{structured scientific spectrum}: it is nonnegative over energy, obeys strict conservation laws, contains sharp discontinuities and singular features, and encodes multi-scale information arising from atomic composition, crystal geometry, and long-range periodicity. These properties make eDOS prediction a challenging learning problem that goes beyond unconstrained regression, motivating models that reason over energy-conditioned structure while preserving physical validity.

\subsection{Phonon Density of States (phDOS)}
\label{app:phdos}

The \emph{phonon density of states} (phDOS) is the lattice-vibrational analog of the eDOS: it characterizes how vibrational normal modes of a crystal are distributed in frequency~\citep{chen2021direct}. For a periodic system with phonon dispersion branches $\{\omega_n(\mathbf{q})\}$, indexed by branch $n$ and crystal momentum $\mathbf{q}$ in the Brillouin zone, the phDOS is defined as
\begin{equation}
g(\omega)
\;=\;
\frac{1}{V_{\mathrm{BZ}}}
\sum_{n}
\int_{\mathrm{BZ}}
\delta\!\left(\omega - \omega_n(\mathbf{q})\right)
\, d\mathbf{q},
\label{eq:phdos_def}
\end{equation}
so that $g(\omega)\,d\omega$ counts the number of vibrational modes with frequency in $[\omega, \omega+d\omega]$. For a unit cell containing $N_a$ atoms, there are exactly $3N_a$ phonon branches: three acoustic branches with $\omega_n(\mathbf{q}) \to 0$ as $\mathbf{q} \to \mathbf{0}$, and $3N_a - 3$ optical branches with finite zone-center frequencies. The mode-counting sum rule
\begin{equation}
\int_{0}^{\omega_{\max}} g(\omega)\, d\omega \;=\; 3 N_a
\label{eq:phdos_mass}
\end{equation}
plays the same role for phDOS that total state counting plays for eDOS, providing an exact conservation constraint that any prediction must satisfy after appropriate normalization.

The phDOS underlies a wide range of thermodynamic, thermal-transport, and dynamic properties of crystalline materials~\citep{hook2013solid}. The lattice contribution to the heat capacity, the vibrational entropy, the Helmholtz free energy, the Debye--Waller factors that govern diffraction intensities, and the temperature dependence of the lattice parameter can all be expressed as integrals of $g(\omega)$ weighted by appropriate Bose--Einstein occupation kernels. Low-frequency behavior in particular is decisive: in the long-wavelength limit, the acoustic branches yield a Debye-like $g(\omega)\propto \omega^{2}$ scaling that controls low-temperature thermodynamics, while the placement of optical features at higher frequencies carries chemical and bonding information about the constituent species.

Structurally, the phDOS exhibits geometric features closely paralleling those of the eDOS. Van Hove singularities arise at critical points of the phonon dispersion, where $\nabla_{\mathbf{q}}\omega_n(\mathbf{q}) = \mathbf{0}$, and manifest as sharp peaks in $g(\omega)$. Gaps between acoustic and optical manifolds, or between distinct groups of optical branches, separate disjoint vibrational windows and are often pronounced in crystals with large mass contrasts among constituent atoms. Localized vibrations associated with light impurities or hydrogen-containing species can produce narrow, isolated bands at high frequency. As with eDOS, the phDOS admits an atom-projected decomposition that quantifies each atomic species' or crystallographic site's contribution to the modes at every frequency, linking global spectral features to local chemistry and providing a natural target for atom-to-frequency attention.

In practice, the phDOS is obtained from density functional perturbation theory (DFPT) or finite-displacement supercell calculations, sampled on a finite frequency grid (e.g.,\ $[0, 1000]$\,cm$^{-1}$), and Gaussian-smeared and sum-normalized for downstream modeling. Predictive models must therefore respect nonnegativity, the mode-counting sum rule, the smooth low-frequency Debye onset, and the sharp Van Hove peaks and inter-branch gaps that determine downstream thermal observables.

\subsection{X-ray Absorption Near-Edge Structure (XANES)}
\label{app:xanes}

\emph{X-ray Absorption Near-Edge Structure} (XANES) describes the X-ray absorption coefficient $\mu(E)$ of a material in a narrow energy window around an inner-shell ionization edge of a chosen \emph{absorbing} atom~\citep{rehr2009ab,mathew2018xas}. Within Fermi's golden rule and in the single-particle picture, the absorption cross section near an edge takes the form
\begin{equation}
\mu(E) \;\propto\; \sum_{f}
\left| \langle f \,|\, \hat{O} \,|\, i \rangle \right|^{2}
\, \delta\!\left(E - (E_f - E_i)\right),
\label{eq:xanes_fgr}
\end{equation}
where $|i\rangle$ is a deep, atom-localized core state on the absorber, $\{|f\rangle\}$ are empty quasiparticle states above the Fermi level, and $\hat{O}$ is the photon--matter interaction operator (dominantly dipolar, with weaker quadrupolar contributions). For K-edge XANES, $|i\rangle$ is the $1s$ orbital of the absorber, dipole-allowed transitions probe local $p$-symmetry projected unoccupied densities at the absorbing site, and the spectrum is referenced to the per-element K-edge energy $E_0$.

A defining property of XANES, distinguishing it sharply from eDOS and phDOS, is that the spectrum is \emph{element- and site-specific}: a single crystal yields a distinct XANES spectrum for each absorbing element it contains (and, in principle, for each crystallographically inequivalent site). Consequently, the same crystal graph may map to several different physically valid spectra depending on the choice of absorber, and predictive models must accept an absorber identity as an additional conditioning input.

Physically, XANES is exquisitely sensitive to the local chemical environment of the absorber. The edge position encodes the absorber's effective charge and oxidation state; pre-edge features carry information about dipole-forbidden or symmetry-relaxed transitions to bound states, with intensities controlled by local site symmetry (e.g., centrosymmetric vs.\ non-centrosymmetric coordination); the prominent post-edge ``white-line'' peak reflects a dense manifold of unoccupied states with appropriate symmetry; and the higher-energy post-edge oscillations arise from multiple scattering of the outgoing photoelectron off neighboring atoms, encoding nearest- and next-nearest-neighbor distances and coordination geometry. A canonical XANES spectrum therefore decomposes into (i) a pre-edge region, (ii) a rising main edge, (iii) one or more white-line peaks immediately above the edge, and (iv) lower-amplitude post-edge oscillations.

After background subtraction and edge-step normalization, XANES is treated as a nonnegative spectrum on a finite, absorber-referenced energy grid (in our experiments, a $128$-bin grid over $[-5, +40]$\,eV relative to $E_0$, with Gaussian smearing and sum normalization). Capturing XANES therefore demands that learned models be absorber-conditioned, that they preserve nonnegativity and sum-normalization, and that they faithfully resolve the relatively sharp edge onset, the high-amplitude white-line, and the lower-amplitude post-edge structure simultaneously, since each region encodes complementary chemical information.

\subsection{Raman Spectra}
\label{app:raman}

\emph{Raman spectra} arise from inelastic light scattering off lattice vibrations: an incident photon of energy $\hbar\omega_{\mathrm{in}}$ scatters off the crystal with outgoing energy $\hbar\omega_{\mathrm{out}}$, exchanging the \emph{Raman shift} $\omega = \omega_{\mathrm{in}} - \omega_{\mathrm{out}}$ with a vibrational mode of the system~\citep{bagheri2023raman}. To leading order in the photon--phonon coupling, the differential scattering cross section is governed by the Placzek polarizability theory~\citep{placzek1934rayleigh}: each Raman-active mode at frequency $\omega_\nu$ contributes a Stokes intensity proportional to the squared derivative of the macroscopic polarizability tensor $\alpha$ with respect to its normal coordinate $Q_\nu$,
\begin{equation}
I_\nu \;\propto\; \frac{\bigl(n(\omega_\nu, T) + 1\bigr)}{\omega_\nu}
\left| \frac{\partial \alpha}{\partial Q_\nu} \right|^{2},
\label{eq:raman_placzek}
\end{equation}
where $n(\omega_\nu, T) = (e^{\hbar \omega_\nu / k_B T} - 1)^{-1}$ is the Bose--Einstein population factor and additional tensor contractions encode the polarization geometry of the incident and scattered photons. For polycrystalline (powder) samples, the observed intensity follows from orientational averaging of the polarizability-derivative tensors over all crystal orientations.

A crucial qualitative feature of Raman spectra is that \emph{symmetry-based selection rules} render only a subset of phonon modes spectroscopically active. In centrosymmetric crystals, the mutual-exclusion rule forbids modes that are simultaneously infrared-active from being Raman-active, and (depending on space group) entire mode irreps may be Raman-silent; for non-centrosymmetric structures, a richer set of modes contributes. Consequently, the Raman spectrum is intrinsically \emph{sparse}: a small number of narrow lines at the active mode frequencies on an otherwise near-zero baseline, with intensities spanning several orders of magnitude across modes.

For modeling purposes, the per-mode intensities $\{(\omega_\nu, I_\nu)\}$ are placed on a finite wavenumber grid (in our experiments, $128$ bins over $[0, 1500]$\,cm$^{-1}$), broadened by a fixed Gaussian kernel ($\sigma = 10$\,cm$^{-1}$) to mimic instrumental and lifetime widths, and sum-normalized to a unit total intensity:
\begin{equation}
I(\omega) \;=\; \frac{1}{Z}\sum_{\nu} I_\nu \, \mathcal{N}(\omega; \omega_\nu, \sigma^2), \qquad
\int I(\omega)\, d\omega = 1.
\end{equation}
Predictive models must respect this sparse-plus-peak geometry: they must localize each active mode to the correct wavenumber, reproduce relative intensities that can vary across several decades, and avoid placing spurious mass in dark regions of the spectrum where selection rules forbid emission. Together with the absence of a useful uniform background and the strong dependence of activity on crystal symmetry, these properties make Raman the most geometrically constrained of the four spectra considered here, and a particularly stringent test of constraint-preserving spectrum prediction.

\section{Constraint-Preserving Physical Reasoning: Mass-Conserving Iterative Refinement}
\label{app:cppr}

This appendix provides a formal, closed-loop derivation of the constraint-preserving physical reasoning stage (``mass-conserving iterative refinement'') introduced in \Cref{sec:reason}. Throughout, let the energy domain be $\Omega=[a,b]\subset\mathbb{R}$, and let $p(\cdot,t)$ denote a time-dependent normalized eDOS density.

\subsection{Transport Model and Conservation Laws}
\label{app:cppr:transport}

\begin{assumption}[Regularity and boundary condition]
\label{asm:regularity}
Assume $p(\cdot,t)$ is integrable on $\Omega$ for each $t$ and $f(\cdot,t)$ is measurable with $j(x,t):=p(x,t)f(x,t)$ well-defined. On bounded $\Omega=[a,b]$, we impose the \emph{no-flux} boundary condition
\begin{equation}
j(a,t)=j(b,t)=0,\quad \forall t.
\label{eq:noflux}
\end{equation}
(For $\Omega=\mathbb{R}$, replace \Cref{eq:noflux} by $\lim_{x\to\pm\infty} j(x,t)=0$ with integrable tails.)
\end{assumption}

\begin{definition}[Continuity equation]
\label{def:continuity}
Given a velocity field $f:\Omega\times[0,1]\to\mathbb{R}$, we define refinement as \emph{pure transport}:
\begin{equation}
\partial_t p(x,t) + \partial_x j(x,t)=0,
\qquad
j(x,t):=p(x,t)f(x,t).
\label{eq:app-continuity}
\end{equation}
\end{definition}

\begin{lemma}[Mass conservation]
\label{lem:mass_conservation}
If $p$ satisfies \Cref{eq:app-continuity} on $\Omega=[a,b]$ and the no-flux condition \Cref{eq:noflux}, then
\begin{equation}
\int_a^b p(x,t)\,dx = \int_a^b p(x,0)\,dx,\quad \forall t.
\label{eq:mass_conserved}
\end{equation}
\end{lemma}
\begin{proof}
Integrate \Cref{eq:app-continuity} over $[a,b]$:
$
\frac{d}{dt}\int_a^b p(x,t)\,dx
= -\int_a^b \partial_x j(x,t)\,dx
= -(j(b,t)-j(a,t)).
$
Apply \Cref{eq:noflux}. \qedhere
\end{proof}

\begin{lemma}[Log-density dynamics]
\label{lem:log_density}
Assume $p(x,t)>0$ at $(x,t)$. Expanding \Cref{eq:app-continuity} yields
\begin{equation}
\partial_t \log p(x,t) = - f(x,t)\,\partial_x \log p(x,t) - \partial_x f(x,t).
\label{eq:log_density}
\end{equation}
Equivalently, along characteristics $x'(t)=f(x(t),t)$,
\begin{equation}
\frac{D}{Dt}\log p(x(t),t) := \partial_t\log p + f\,\partial_x\log p
= -\partial_x f.
\label{eq:material_log_density}
\end{equation}
\end{lemma}
\begin{proof}
By product rule, \Cref{eq:app-continuity} is
$
\partial_t p + f\,\partial_x p + p\,\partial_x f = 0.
$
Divide by $p>0$ to obtain \Cref{eq:log_density}. Substitute $x'(t)=f$ to get \Cref{eq:material_log_density}. \qedhere
\end{proof}

\subsection{Wasserstein Displacement Interpolation as Reference Flow}
\label{app:cppr:ot}

Let $p_0$ be the initial normalized prediction and $p_T$ the target normalized density. Define CDFs
$
F_0(x)=\int_{-\infty}^x p_0(u)\,du,\
F_T(x)=\int_{-\infty}^x p_T(u)\,du,
$
and quantiles (generalized inverses)
$
Q_0(s)=F_0^{-1}(s),\
Q_T(s)=F_T^{-1}(s),\ s\in[0,1].
$

\begin{lemma}[1D optimal transport map]
\label{lem:1d_ot_map}
If $p_0$ and $p_T$ are absolutely continuous on $\Omega$, the unique optimal transport map for the quadratic cost pushes $p_0$ to $p_T$ and is given by the monotone rearrangement
\begin{equation}
T(x)=F_T^{-1}(F_0(x)).
\label{eq:ot_map}
\end{equation}
\end{lemma}
\begin{proof}
This is a standard 1D result: the optimal coupling is induced by matching equal cumulative mass, i.e., $F_0(x)=F_T(T(x))$, hence $T(x)=F_T^{-1}(F_0(x))$. \qedhere
\end{proof}

\begin{definition}[Displacement interpolation]
\label{def:displacement}
Define the Lagrangian particle path for mass label $s\in[0,1]$:
\begin{equation}
x_t(s) := (1-t)Q_0(s) + t Q_T(s),\qquad t\in[0,1].
\label{eq:disp_interp_path}
\end{equation}
Let $p_t$ be the pushforward of $p_0$ under $x_t(\cdot)$, i.e., $p_t = (x_t)_\# p_0$.
\end{definition}

\begin{lemma}[Closed-form density along the geodesic]
\label{lem:geodesic_density}
Assume $Q_0,Q_T$ are differentiable a.e. Then for $x=x_t(s)$,
\begin{equation}
p_t(x_t(s))=\frac{1}{\partial_s x_t(s)}
=\Big((1-t)\,\partial_s Q_0(s) + t\,\partial_s Q_T(s)\Big)^{-1}.
\label{eq:pt_quantile}
\end{equation}
\end{lemma}
\begin{proof}
In 1D, pushing forward the uniform measure on $s\in[0,1]$ through $x_t(\cdot)$ yields density $p_t(x)=\frac{ds}{dx} = \frac{1}{\partial_s x_t(s)}$ at $x=x_t(s)$. \qedhere
\end{proof}

\begin{lemma}[Reference velocity field and continuity equation]
\label{lem:ref_velocity_continuity}
Define the reference velocity in Lagrangian coordinates (equivalently, the reference field $f_t(x)$ from the main text):
\begin{equation}
f^\star(x_t(s),t) := \partial_t x_t(s) = Q_T(s)-Q_0(s).
\label{eq:ref_velocity}
\end{equation}
Then the pair $(p_t,f^\star)$ satisfies the continuity equation \Cref{eq:app-continuity} on $t\in[0,1]$.
\end{lemma}
\begin{proof}
Each label $s$ moves along the trajectory $x_t(s)$ with velocity $\partial_t x_t(s)$. The induced Eulerian velocity field $f^\star(\cdot,t)$ transports mass with no creation/destruction; by standard change-of-variables for pushforwards, $(p_t,f^\star)$ solves \Cref{eq:app-continuity}. \qedhere
\end{proof}

\subsection{Learning the Velocity Field (Simulation-Free Regression)}
\label{app:cppr:learning}

We learn a neural approximation $f_\theta(x,t;C)$ (optionally conditioned on context $C$ identifying each crystal) by regressing to the reference field $f^\star$:
\begin{equation}
\mathcal{L}(\theta) = \mathbb{E}_{(x,t)\sim \mathcal{D}}\big[\|f_\theta(x,t;C)-f^\star(x,t)\|_2^2\big].
\label{eq:velocity_regression}
\end{equation}
Because $f^\star$ is constructed from the OT geodesic and satisfies \Cref{lem:ref_velocity_continuity}, minimizing \Cref{eq:velocity_regression} encourages $f_\theta$ to approximate a physically valid, mass-conserving transport operator.

\subsection{CDF Transport and Equivalence to Density Transport}
\label{app:cppr:cdf}

Define the CDF of $p(\cdot,t)$ on $\mathbb{R}$ (or on $\Omega$ with appropriate extension)
\begin{equation}
F(x,t):=\int_{-\infty}^x p(u,t)\,du.
\label{eq:cdf_def}
\end{equation}

\begin{lemma}[CDF advection PDE]
\label{lem:cdf_pde}
If $p$ satisfies \Cref{eq:app-continuity} with velocity $f$, then $F$ satisfies the first-order advection equation
\begin{equation}
\partial_t F(x,t) + f(x,t)\,\partial_x F(x,t)=0.
\label{eq:cdf_pde}
\end{equation}
Conversely, if $F$ solves \Cref{eq:cdf_pde} and is differentiable in $x$, then $p=\partial_x F$ solves \Cref{eq:app-continuity}.
\end{lemma}
\begin{proof}
Differentiate \Cref{eq:cdf_def} in $t$ and use \Cref{eq:app-continuity}:
$
\partial_t F(x,t)=\int_{-\infty}^x \partial_t p\,du
= -\int_{-\infty}^x \partial_u(p f)\,du
= -(p f)(x,t).
$
Since $\partial_x F=p$, we obtain $\partial_t F + f\,\partial_x F=0$.
Conversely, differentiating \Cref{eq:cdf_pde} w.r.t.\ $x$ yields \Cref{eq:app-continuity}. \qedhere
\end{proof}

\subsection{Discrete Upwind Scheme and Discrete Validity}
\label{app:cppr:discretization}

Let $\{x_i\}_{i=0}^{N_x-1}$ be a uniform grid on $\Omega$ with spacing $\Delta x$, and $\{t_k\}_{k=0}^{N_t-1}$ with step $\Delta t$. Let $F_i^k\approx F(x_i,t_k)$ and $f_i^k:=f_\theta(x_i,t_k)$.

\begin{definition}[Upwind update]
\label{def:upwind}
We update the CDF by a forward-Euler, first-order upwind scheme:
\begin{equation}
F_i^{k+1} = F_i^k - \Delta t\, f_i^k\, D_x^{\mathrm{up}}F_i^k,
\label{eq:upwind}
\end{equation}
where
\[
D_x^{\mathrm{up}}F_i^k=
\begin{cases}
\frac{F_i^k-F_{i-1}^k}{\Delta x}, & f_i^k\ge 0,\\[6pt]
\frac{F_{i+1}^k-F_i^k}{\Delta x}, & f_i^k< 0.
\end{cases}
\]
\end{definition}

\begin{lemma}[Discrete monotonicity and nonnegativity]
\label{lem:discrete_monotone}
Assume a Courant–Friedrichs–Lewy (CFL) condition $\max_i |f_i^k|\frac{\Delta t}{\Delta x}\le 1$ for all $k$. If $F^k$ is nondecreasing in $i$ (i.e., $F_{i}^k\ge F_{i-1}^k$), then $F^{k+1}$ is also nondecreasing. Consequently, the recovered density
\begin{equation}
p_i^{k}:=\frac{F_i^{k}-F_{i-1}^{k}}{\Delta x}\ge 0
\label{eq:density_from_cdf}
\end{equation}
is nonnegative for all $i$.
\end{lemma}
\begin{proof}
Under upwinding and CFL, \Cref{eq:upwind} is a convex combination of neighboring $F$ values; monotonicity of $F$ is preserved by monotone schemes. Then \Cref{eq:density_from_cdf} is a finite difference of a nondecreasing sequence, hence nonnegative. \qedhere
\end{proof}

\begin{lemma}[Discrete mass conservation (up to boundary/CFL error)]
\label{lem:discrete_mass}
Let $p_i^k$ be defined by \Cref{eq:density_from_cdf}. If $F_0^k\approx 0$ and $F_{N_x-1}^k\approx 1$ are fixed (no-flux/consistent boundary handling), then
\[
\sum_{i=1}^{N_x-1} p_i^k\,\Delta x = F_{N_x-1}^k - F_0^k \approx 1
\]
for all $k$, i.e., total mass is preserved up to numerical precision. A final renormalization may be applied to correct rounding.
\end{lemma}
\begin{proof}
By telescoping,
$
\sum_{i=1}^{N_x-1} p_i^k\,\Delta x
=\sum_{i=1}^{N_x-1}(F_i^k-F_{i-1}^k)=F_{N_x-1}^k-F_0^k.
$
Boundary consistency yields $\approx 1$. \qedhere
\end{proof}

\subsection{End-to-End Validity Guarantee}
\label{app:cppr:closure}

\begin{theorem}[Constraint-preserving refinement]
\label{thm:constraint_preserving}
Assume the refinement evolves a valid initial CDF $F(\cdot,0)$ (nondecreasing with limits $0$ and $1$) under \Cref{lem:cdf_pde} in the continuous setting, or under \Cref{eq:upwind} with the CFL condition in the discrete setting. Then the resulting refined density $p(\cdot,t)=\partial_x F(\cdot,t)$ (or $p_i^k$ via \Cref{eq:density_from_cdf}) satisfies:
\begin{enumerate}
\item \textbf{Nonnegativity:} $p(\cdot,t)\ge 0$ (or $p_i^k\ge 0$);
\item \textbf{Mass conservation:} $\int_\Omega p(\cdot,t)\,dx=1$ (or $\sum_i p_i^k\Delta x\approx 1$);
\item \textbf{Pure transport:} refinement redistributes spectral mass without creation/destruction.
\end{enumerate}
\end{theorem}
\begin{proof}
Continuous case: by \Cref{lem:cdf_pde}, density transport is equivalent to CDF advection; \Cref{lem:mass_conservation} gives conservation, and CDF monotonicity implies nonnegativity. Discrete case: \Cref{lem:discrete_monotone} gives nonnegativity and \Cref{lem:discrete_mass} gives conservation. \qedhere
\end{proof}

\paragraph{Remark.}
The OT reference field $f^\star$ in \Cref{lem:ref_velocity_continuity} is \emph{itself} a solution of the continuity equation, so regressing $f_\theta\approx f^\star$ aligns learning with physically valid transport. Inference then solves the transport PDE in a form (\Cref{lem:cdf_pde}) and scheme (\Cref{lem:discrete_monotone,lem:discrete_mass}) that preserve constraints by construction, yielding an end-to-end closed-loop refinement module.

\section{Algorithm}
\label{app:algo}

This section summarizes the training and inference pipeline of DeepSciReasoner, including (i) structure-aware spectrum decoding and (ii) constraint-preserving physical reasoning via mass-conserving iterative refinement.

\paragraph{Notation.}
Let a crystal be represented as a graph $G=(V,E)$ with node features and edge features. Let the discretized energy grid be $\{x_i\}_{i=0}^{N_x-1}\subset[a,b]$ with spacing $\Delta x$. The ground-truth (unnormalized) eDOS is $y\in\mathbb{R}_+^{N_x}$ and its normalized density is
$p_T = y / \sum_i y_i$.
The decoder predicts $\hat{y}_0\in\mathbb{R}_+^{N_x}$ and its normalized density is
$p_0 = \hat{y}_0 / \sum_i \hat{y}_{0,i}$.
Let $F_0$ denote the CDF of $p_0$.
The refinement velocity network is $f_\theta(x,t;C)$, optionally conditioned on context $C$ (e.g., crystal embedding).

\subsection{Training and Inference Procedure}
\label{sec:pipeline:algo}

We show the training and inference pipelines of DeepSciReasoner in \Cref{alg:deepscireasoner}. We also provide some useful notes for practical implementation below.

\begin{algorithm}[ht]
\caption{(DeepSciReasoner) Training and Inference}
\label{alg:deepscireasoner}
\begin{algorithmic}[1]
\Require Training set $\mathcal{D}=\{(G^{(n)},y^{(n)})\}_{n=1}^N$; energy grid $\{x_i\}_{i=0}^{N_x-1}$; refinement horizon $T$; number of time grid points $N_t$; $\Delta t=T/(N_t-1)$
\Require Decoder parameters $\phi$; refinement parameters $\theta$

\vspace{2pt}
\Statex \texttt{(Training) Structure-aware Spectrum Decoding}
\For{each minibatch}
    \State Encode crystal with edge-aware crystal graph Transformer: $\{h_i\},\,h_x \gets \mathrm{Enc}_\phi(G)$
    \State Decode eDOS with energy self-attention + source (atom-to-energy) attention:
    \State \hspace{1.1em} $\hat{y}_0 \gets \mathrm{Dec}_\phi(\{h_i\},h_x,\{x_i\})$
    \State Normalize: $p_T \gets y/\sum_i y_i$, \quad $p_0 \gets \hat{y}_0/\sum_i \hat{y}_{0,i}$
    \State Compute reconstruction loss (distributional + mass/head as desired), e.g.:
    \State \hspace{1.1em} $\mathcal{L}_{\mathrm{dec}}(\phi)\gets D_{\mathrm{KL}}(p_T\|p_0) + \lambda\,\|\widehat{Z}-Z\|_2^2$
    \State Update decoder parameters: $\phi \leftarrow \phi - \eta_\phi \nabla_\phi \mathcal{L}_{\mathrm{dec}}$
\EndFor

\vspace{2pt}
\Statex \texttt{(Training) Mass-Conserving Iterative Refinement}
\For{each minibatch $\{(G,y)\}\sim \mathcal{D}$}
    \State \textit{(Freeze $\phi$)} Compute initial prediction $\hat{y}_0$ via Stage I; normalize to $p_0$
    \State Normalize ground truth to $p_T \gets y/\sum_i y_i$
    \State Build OT displacement-interpolation reference velocity field $f^\star(x,t)$:
    \State \hspace{1.1em} Compute CDFs $F_0,F_T$ and quantiles $Q_0,Q_T$; define $x_t(s)=(1-t)Q_0(s)+tQ_T(s)$
    \State \hspace{1.1em} Set $f^\star(x_t(s),t)=Q_T(s)-Q_0(s)$ and convert to Eulerian form (via $s=F_t(x)$)
    \State Sample training pairs $(x,t)\sim \mathrm{Unif}([a,b]\times[0,1])$ (or from $p_t$) and context $C\gets h_x$
    \State Velocity regression loss: $\mathcal{L}_{\mathrm{vel}}(\theta)\gets \mathbb{E}_{(x,t)}\|f_\theta(x,t;C)-f^\star(x,t)\|_2^2$
    \State Update refinement parameters: $\theta \leftarrow \theta - \eta_\theta \nabla_\theta \mathcal{L}_{\mathrm{vel}}$
\EndFor

\vspace{2pt}
\Statex \texttt{(Inference) Decoding $\rightarrow$ Reasoning}
\Function{Infer}{$G$}
    \State $\{h_i\},\,h_x \gets \mathrm{Enc}_\phi(G)$; \quad $\hat{y}_0 \gets \mathrm{Dec}_\phi(\{h_i\},h_x,\{x_i\})$
    \State $p_0 \gets \hat{y}_0/\sum_i \hat{y}_{0,i}$; \quad $F^0_i \gets \sum_{m\le i} p_{0,m}\Delta x$ \Comment{initial CDF}
    \For{$k=0,1,\dots,N_t-2$}
        \State $t_k \gets k\Delta t$; \quad $C\gets h_x$
        \State Evaluate velocity on grid: $v_i \gets f_\theta(x_i,t_k;C)$
        \State Upwind CDF update (advection): $F^{k+1}_i \gets F^k_i - \Delta t \, v_i \, D_x^{\mathrm{up}}F^k_i$
        \State Recover density: $p^{k+1}_i \gets (F^{k+1}_i - F^{k+1}_{i-1})/\Delta x$; renormalize $p^{k+1} \leftarrow p^{k+1}/\sum_i p^{k+1}_i\Delta x$
    \EndFor
    \State \Return final density $p^{N_t-1}$; optionally output $\hat{y} \gets p^{N_t-1}\cdot \widehat{Z}$.
\EndFunction
\end{algorithmic}
\end{algorithm}

\paragraph{Reference Trajectory.}
The OT reference trajectory $\mathrm{OTVelocity}(p_0,p_T,\{x_i\})$ computes the 1D $W_2$ displacement-interpolation velocity by (i) forming CDFs on the grid, (ii) computing quantiles via inverse-CDF interpolation, (iii) defining $x_t(s)=(1-t)Q_0(s)+tQ_T(s)$ and $f^\star(x_t(s),t)=Q_T(s)-Q_0(s)$, and (iv) converting to Eulerian coordinates via $s=F_t(x)$ (linear interpolation on the grid).

\paragraph{Upwind Derivative.}
For $D_x^{\mathrm{up}}$, use
\[
D_x^{\mathrm{up}}(F)_i =
\begin{cases}
(F_i-F_{i-1})/\Delta x, & v_i \ge 0,\\
(F_{i+1}-F_i)/\Delta x, & v_i < 0,
\end{cases}
\]
with boundary handling consistent with no-flux (or fixed-endpoint) conditions.

\paragraph{Practical Stability.}
Enforce a CFL-like condition $\max_i |v_i|\,\Delta t/\Delta x \le 1$ (via smaller $\Delta t$ or clamping) to keep the CDF monotone and the density nonnegative.

\section{Additional Experimental Setup Details}
\label{app:hyper}

\subsection{Data Curation Setup}\label{app:data}

\paragraph{eDOS.}
We evaluate on two benchmark datasets, CalTech~\citep{Kong2022} and TRI~\citep{mueller2024electrochemical}, both derived from Materials Project structures and DFT electronic structure calculations~\citep{Jain2013}. However, CalTech includes some unstable crystal structures while TRI only includes strictly stable crystal structures. The target is the electronic density of states (eDOS) ($D(E)$), which describes how densely electronic states populate each energy level ($E$). A central feature of realistic spectra is the band gap: for an insulator/semiconductor, there exists an energy interval with (near-)zero density separating the valence band (occupied states) from the conduction band (unoccupied states). The valence-band maximum (VBM) and conduction-band minimum (CBM) define the gap edges, with ($E_g = E_{\mathrm{CBM}} - E_{\mathrm{VBM}} \geq 0$). Unlike previous work that removes the interval ($[E_{\mathrm{VBM}}, E_{\mathrm{CBM}}]$) for gapped materials---effectively ``splicing'' the spectrum and sidestepping gap prediction---we retain the full eDOS so the model must learn both (i) to place the gap at the correct energy and (ii) to maintain the sharp spectral transition at the band edges. This design is closer to the physical object returned by electronic-structure calculations and makes evaluation more stringent, since small errors near ($E_{\mathrm{VBM}})$ and ($E_{\mathrm{CBM}}$) can blur the gap or introduce spurious in-gap states.

Both eDOS datasets apply smoothing to reduce discretization noise and broaden extremely sharp features that arise from finite ($k$)-point sampling. CalTech applies a binning procedure to the raw DOS, where the points within each bin are averaged and assigned to the energy corresponding to the bin center. In contrast, TRI uses Gaussian smoothing ($\sigma=2$~eV), equivalent to convolving the spectrum with a Gaussian kernel, which more aggressively suppresses high-frequency fluctuations but can also broaden narrow peaks and soften steep band edges. After smoothing, the datasets discretize the spectrum on uniform energy grids: CalTech uses 128 bins over $[-4,4]$~eV, while TRI uses 256 bins over $[-5,5]$~eV. The differing bin counts and energy ranges imply different effective energy resolutions and tail coverage, which can affect the apparent sharpness of peaks and the numerical width of gaps. Overall, retaining the full spectrum---together with dataset-specific smoothing and discretization---yields a challenging, physically faithful benchmark that emphasizes band-edge fidelity, gap sharpness, and the model's ability to reproduce fine-grained spectral structure rather than only coarse integrated properties.

\paragraph{phDOS.}
We use the curated DFPT phonon DOS dataset of~\citet{chen2021direct}, which contains $1{,}521$ crystalline materials drawn from Materials Project and computed with density functional perturbation theory (DFPT). The target is the phonon density of states $g(\omega)$, which counts vibrational normal modes per unit frequency; integration over $g(\omega)$ recovers the total number of modes ($3 N_a$ for a unit cell with $N_a$ atoms), making mode-counting an exact sum rule that any predictive model must respect after normalization. Realistic phonon spectra are physically structured: a smooth Debye-like onset at low frequency from the acoustic branches, sharp Van Hove peaks at higher frequencies from critical points in the optical branches, and---when constituent atoms differ substantially in mass---pronounced gaps separating acoustic and optical manifolds. Unlike previous work that focused only on coarse summaries such as low-frequency moments used for heat-capacity prediction, we retain the full $[0,1000]$~cm$^{-1}$ window on the 51-bin grid distributed with the dataset and sum-normalize the bin values to integral one for distributional training. This full-spectrum setting forces the model to learn both the smooth long-wavelength acoustic behavior and the sparse optical peak structure simultaneously, emphasizing faithful reproduction of mode counting, peak placement, and inter-branch gaps rather than only coarse integrated quantities like the lattice heat capacity.

\paragraph{XANES.}
We curate K-edge spectra from the Materials Project XAS database~\citep{Jain2013,mathew2018xas}, which holds $157{,}420$ K-edge spectra across all MP materials; intersected with our $37{,}565$-mpid crystal graph corpus, this yields $50{,}693$ candidate (mpid, absorber) pairs. The target is the X-ray absorption coefficient $\mu(E)$ in a narrow energy window around the K-edge of a chosen absorbing element, which probes element-specific empty $p$-symmetry states above the Fermi level (with weaker quadrupole contributions). Unlike eDOS and phDOS, the same crystal yields a distinct XANES spectrum for each absorbing element it contains; consequently, the decoder must accept the absorber identity as an additional conditioning input, and we extend the model with a learned embedding over the absorber atomic number that is injected into the crystal context so that the same crystal graph can produce different spectra for different absorbers.

We convert each spectrum to a relative-energy axis $E_{\mathrm{rel}} = E - E_0$, where $E_0$ is the per-element K-edge energy returned by Materials Project, and interpolate it onto a uniform 128-bin grid over $[-5, +40]$~eV. The window is chosen empirically so that the rising edge, the white-line peak, and lower-amplitude post-edge oscillations are covered for $\geq 98\%$ of K-edge absorbers (MP XAS spans roughly $[-2,+52]$~eV for $s/p$ absorbers and $[-17,+37]$~eV for $3d$/heavier $d$-block elements), and we apply flat extrapolation (zero below the edge, plateau above) for spectra whose raw axis is narrower than the window rather than rejecting them outright. We then apply Gaussian smearing with $\sigma = 0.5$~eV (matched to the $\sim 0.35$~eV bin spacing) and sum-normalize to integral one. We reject a (mpid, absorber) pair if (a) its K-edge energy $E_0$ is invalid, (b) its raw axis fails to overlap $[-5,+40]$~eV at all, or (c) its post-smear spectrum has non-positive integral; $798$ of $50{,}693$ candidate pairs are gated out, leaving $49{,}895$ valid records. To enable apples-to-apples comparison with the one-spectrum-per-material convention adopted for eDOS, phDOS, and Raman, we additionally construct an oxygen K-edge subset by post-hoc filtering the full manifest to records whose absorber is oxygen, yielding $5{,}094$ mpids on which baselines built around the one-spectrum-per-material convention can be compared on equal footing. As with eDOS, retaining the full $[-5,+40]$~eV window---rather than masking the pre-edge or post-edge region---yields a stringent benchmark that emphasizes edge-position accuracy, white-line shape, and post-edge oscillation structure, since small errors near the edge can shift the apparent oxidation state and distort downstream chemical interpretation.

\paragraph{Raman.}
We use the high-throughput DFPT Raman database of~\citet{bagheri2023raman}, distributed via the Materials Cloud Archive (DOI 10.24435/materialscloud:ze-58) as a manifest of $5{,}099$ DFPT-computed Raman tensor records indexed by mpid. The target is the powder-averaged Stokes Raman intensity as a function of Raman shift, set by symmetry-allowed phonon modes and their polarizability derivatives, and so intrinsically \emph{sparse}: a handful of narrow lines on an otherwise near-zero baseline, with per-mode intensities spanning several orders of magnitude. For each Raman-active mode we compute the powder-averaged Raman activity from the corresponding $3\times 3$ DFPT Raman tensor $R$ via the Placzek invariants~\citep{placzek1934rayleigh} (i.e., the Stokes intensity of \Cref{eq:raman_placzek} up to the temperature- and frequency-dependent Bose prefactor),
\begin{equation*}
a' = \tfrac{1}{3}\,\mathrm{tr}(R), \quad
g'^{\,2} = \tfrac{1}{2}\sum_{i<j}(R_{ii} - R_{jj})^2 + 3\,(R_{xy}^2 + R_{yz}^2 + R_{zx}^2),
\end{equation*}
with $I_{\mathrm{mode}} \propto 45\,a'^{\,2} + 7\,g'^{\,2}$. We then place a Gaussian of $\sigma = 10$~cm$^{-1}$ at each mode frequency $\omega_\nu$, weighted by $I_{\mathrm{mode}}$, on a uniform 128-bin grid over $[0, 1500]$~cm$^{-1}$ (covering $>99\%$ of inorganic-crystal modes; the bin width $\approx 11.7$~cm$^{-1}$ pairs naturally with $\sigma=10$), and sum-normalize the resulting spectrum to integral one. We reject a record if (a) the frequencies and tensor fields are missing or of unequal length, (b) all per-mode Raman tensors are absent (centrosymmetric crystals have no Raman-active modes and therefore a zero spectrum), or (c) the sum-normalized spectrum integrates to zero; $2{,}816$ of the $5{,}099$ candidate records that intersect our crystal graph corpus are gated out (mostly centrosymmetric), leaving $2{,}283$ valid records. Since only one Raman spectrum exists per material, no absorber conditioning is required and the eDOS encoder is reused without modification. The remaining sparsity-plus-peak geometry makes pointwise regression particularly fragile---a small displacement of mass into a Raman-silent region of the spectrum produces a visually wrong prediction---which makes Raman the most geometrically constrained of our four benchmarks and a particularly stringent test of constraint-preserving spectrum prediction.

\paragraph{Splits.} All datasets use an 80/10/10 train/val/test split with a fixed random seed (\texttt{random\_state=1}): $39{,}916$/$4{,}990$/$4{,}989$ records for the full multi-absorber XANES manifest, $4{,}075$/$510$/$509$ for the oxygen K-edge subset, and $1{,}826$/$229$/$228$ for Raman. The eDOS and phDOS splits follow the conventions of the CalTech/TRI~\citep{Kong2022,mueller2024electrochemical} and~\citet{chen2021direct} releases, respectively.

\subsection{Training Setup}
\paragraph{Edge-Aware Crystal Graph Transformer.}
Our \emph{Edge-Aware Crystal Graph Transformer} encoder comprises three stacked \texttt{TransformerConv} layers~\citep{Fey/Lenssen/2019} with hidden size 512. We set both the node embedding dimension \(d_h\) and edge embedding dimension \(d_e\) to 512, and use 4 attention heads per layer. To regularize the encoder, we apply node-feature dropout with rate \(p=0.2\) in each layer, placed after \texttt{BatchNorm}~\citep{ioffe2015batch} and \texttt{LeakyReLU}.

\paragraph{Structure-to-Spectrum Decoder.}
The \emph{Structure-to-Spectrum Decoder} predicts eDOS on an energy grid using a two-stage decoding design. The first decoder module attends from positional (energy-bin) queries to the crystal graph representations to produce an intermediate spectrum representation; the second module then performs an additional round of attention refinement, conditioned on the same crystal representations. Each module contains 6 Transformer layers (12 layers total). We use an embedding dimension of 512, feed-forward dimension 2048, and 8 attention heads per layer. We adopt a Pre-\texttt{LayerNorm} formulation, where layer normalization precedes the attention and FFN sublayers, which improves optimization stability relative to Post-LayerNorm. The final output is passed through \texttt{Softplus}~\citep{zheng2015improving} to enforce non-negativity, yielding an unnormalized nonnegative eDOS $\hat{y}$; its sum gives the predicted total mass $\hat{Z}=\sum_i\hat{y}_i$, while its sum-normalized form is treated as a discrete probability distribution for the shape (KL) term.

\paragraph{Mass-Conserving Iterative Refinement.}
Starting from the initial prediction, we apply a \emph{Mass-Conserving Iterative Refinement} module that improves the density via a learned transport (velocity) field while preserving total mass. The velocity-field network follows a simplified variant of the decoder architecture and keeps the same core hyperparameters (embedding size, FFN size, and number of heads) to model fine-grained transport dynamics. For refinement, we discretize the time horizon \(t \in [0,T]\) with \(T=1\) and \(N_t=100\) time grid points, giving \(\Delta t = T/(N_t-1) \approx 0.0101\).

\paragraph{Training Details.}
Training proceeds in two stages. In \textbf{Stage I}, we jointly train the structure encoder and spectrum decoder for 200 epochs with batch size 128 using Adam~\citep{kinga2015method} with \(\beta=(0.9,0.98)\) and \(\epsilon=10^{-9}\), following the original Transformer~\citep{vaswani2017attention}; the learning rate of $1.0$ is a base rate scaled by an inverse-square-root warmup schedule. In \textbf{Stage II}, we freeze the pretrained encoder and decoder and train only the refinement (velocity-field) network for 200 epochs with batch size 128 using AdamW~\citep{loshchilov2017decoupled} (learning rate 0.01, weight decay 0.001). We use a \texttt{Cosine} annealing learning-rate schedule with a \texttt{Linear} warmup over the first 10\% of training steps. All modules are implemented in \texttt{PyTorch} and trained on a server with 4 NVIDIA H100 GPUs. \Cref{tab:hyperparams} summarizes all hyperparameters.

\begin{table}[ht!]
\centering
\caption{Hyperparameter settings for DeepSciReasoner.}
\label{tab:hyperparams}
\small
\setlength{\tabcolsep}{6pt}
\begin{tabular}{ll}
\toprule
\textbf{Component / Setting} & \textbf{Value} \\
\midrule
\multicolumn{2}{l}{\textbf{Edge-Aware Crystal Graph Transformer (Encoder)}} \\
\midrule
Backbone operator & \texttt{TransformerConv}~\citep{Fey/Lenssen/2019} \\
\# encoder layers & 3 \\
Hidden size & 512 \\
Node embedding dim \(d_h\) & 512 \\
Edge embedding dim \(d_e\) & 512 \\
\# attention heads / layer & 4 \\
Normalization & \texttt{BatchNorm}~\citep{ioffe2015batch} \\
Activation & \texttt{LeakyReLU} \\
Node feature dropout \(p\) & 0.2 \\
\midrule
\multicolumn{2}{l}{\textbf{Structure-to-Spectrum Decoder}} \\
\midrule
Decoder modules & 2 \\
Layers per module & 6 \\
Total decoder layers & 12 \\
Embedding dim & 512 \\
FFN dim & 2048 \\
\# attention heads / layer & 8 \\
LayerNorm placement & Pre-\texttt{LayerNorm} \\
Output activation & \texttt{Softplus}~\citep{zheng2015improving} \\
Output normalization & \texttt{SumNorm} \\
\midrule
\multicolumn{2}{l}{\textbf{Mass-Conserving Iterative Refinement (MCIR)}} \\
\midrule
Velocity net embedding dim & 512 \\
Velocity net FFN dim & 2048 \\
Velocity net \# heads / layer & 8 \\
Time horizon \(T\) & 1 \\
\# time grid points \(N_t\) & 100 \\
Step size \(\Delta t = T/(N_t-1)\) & \(\approx 0.0101\) \\
\midrule
\multicolumn{2}{l}{\textbf{Stage I: Encoder + Decoder Training}} \\
\midrule
Epochs & 200 \\
Batch size & 128 \\
Optimizer & Adam~\citep{kinga2015method} \\
Learning rate (base, warmup-scaled) & 1.0 \\
Adam \(\beta\) & (0.9, 0.98) \\
Adam \(\epsilon\) & \(10^{-9}\) \\
\midrule
\multicolumn{2}{l}{\textbf{Stage II: MCIR Training (Encoder/Decoder Frozen)}} \\
\midrule
Epochs & 200 \\
Batch size & 128 \\
Optimizer & AdamW~\citep{loshchilov2017decoupled} \\
Learning rate & 0.01 \\
Weight decay & 0.001 \\
LR schedule & \texttt{Cosine} annealing + \texttt{Linear} warmup \\
Warmup ratio & 10\% of total steps \\
\midrule
\multicolumn{2}{l}{\textbf{Implementation}} \\
\midrule
Framework & \texttt{PyTorch} \\
Hardware & 4 $\times$ NVIDIA H100 GPUs \\
\bottomrule
\end{tabular}
\end{table}

\subsection{Baselines}\label{app:base}
We benchmark against representative methods for learning from periodic crystal structures and for predicting structured spectra. Overall, the baselines fall into two families: (i) crystal/graph structure encoders that learn a global representation of a crystal and map it to a target through a regression or spectrum head, and (ii) crystal-to-sequence predictors that explicitly model the spectrum as an ordered sequence over discrete bins. Because eDOS is the only target with a widely adopted neural-baseline suite, structure encoders form the common backbone across all four spectra (eDOS, phDOS, XANES, Raman), while crystal-to-sequence predictors are used only on eDOS, where they were originally proposed. For each encoder we keep the optimizer and message-passing hyperparameters recommended by the baseline's original implementation; for non-eDOS targets we additionally replace the scalar regression head with a sum-pooled spectrum head matching the target grid, followed by \texttt{Softplus} non-negativity and sum normalization, while leaving the encoder architecture unchanged.

\paragraph{Structure Encoders with Regression/Spectrum Head.}
\begin{enumerate}
    \item \textbf{E3NN (E(3)-equivariant neural networks)}~\citep{geiger2022e3nn}. A general framework for building neural networks that are equivariant to 3D Euclidean transformations (rotations, translations, and in some settings reflections), using spherical harmonics and tensor products to ensure features transform predictably under coordinate changes. This inductive bias is well matched to atomistic systems where absolute orientation should not affect predictions. In our benchmarking, E3NN represents the family of geometric equivariant message-passing / tensor-field models commonly used for 3D materials learning. We evaluate E3NN on all four spectrum targets (eDOS, phDOS, XANES, Raman).
    \item \textbf{GATGNN (graph attention GNN with global attention)}~\citep{louis2020graph}. A crystal graph neural network designed for inorganic materials property prediction. It stacks multiple (augmented) graph-attention layers for message passing and includes a global attention mechanism to aggregate node features into a graph-level embedding. Compared to vanilla MPNNs, its core distinction is the attention-based weighting of neighbor contributions and the explicit global attention readout, aimed at improving expressivity and long-range information integration. We evaluate GATGNN on all four spectrum targets.
    \item \textbf{CrystalFormer (conformal crystal graph Transformer)}~\citep{wang2024conformal}. A Transformer-style crystal encoder that emphasizes robust handling of periodicity and geometric information, with (i) a graph construction/encoding strategy that captures periodic invariance and (ii) an efficient way to incorporate angular information; the resulting architecture is a graph Transformer tailored to periodic structures and serves as a strong general-purpose crystal representation learner for downstream regression tasks. We evaluate CrystalFormer on eDOS only, where prior work establishes it as a strong baseline.
    \item \textbf{MEGNet}~\citep{Chen2019}. A universal graph network for molecules and crystals that performs alternating node/edge/global feature updates. We adopt the standard MEGNet block (three message-passing layers, set-to-set readout) and replace the scalar regression head with a sum-pooled spectrum head. We evaluate MEGNet on phDOS, XANES, and Raman, where it serves as a representative of graph-network ensemble baselines used in prior work.
    \item \textbf{M3GNet}~\citep{chen2022m3gnet}. A universal interatomic-potential graph network that includes three-body interactions and a many-body angular term. We use the pretrained M3GNet element embeddings, finetune the message-passing block on each target spectrum, and replace the per-atom energy head with a sum-pooled spectrum head. We evaluate M3GNet on phDOS, XANES, and Raman.
\end{enumerate}

\paragraph{Structured Spectrum Predictors.}
\begin{enumerate}
    \item \textbf{Mat2Spec (probabilistic embedding with contrastive learning for spectra)}~\citep{Kong2022}. A materials-to-spectrum framework introduced for predicting spectral properties such as electronic/phonon DOS. It frames spectrum prediction through probabilistic embeddings and uses contrastive learning to learn representations that support accurate spectrum inference; the work demonstrates utility for tasks such as identifying gaps and screening candidates with follow-up ab initio validation. In our comparisons, Mat2Spec serves as a canonical baseline that treats DOS prediction as a dedicated spectrum-learning problem rather than a generic scalar property regression.
    \item \textbf{DOSTransformer (prompt-guided multi-modal Transformer)}~\citep{lee2023density}. DOSTransformer explicitly models the fact that DOS values depend on both the crystal structure and the energy coordinate. It integrates heterogeneous inputs from the material and from energy levels using a multi-modal Transformer with cross-attention/self-attention, and introduces crystal-system prompts to guide learning of interactions that are shared across (and within) crystal systems. This makes it a strong baseline for directly predicting the DOS sequence over energy bins, rather than predicting a single pooled target.
\end{enumerate}

\paragraph{DeepSciReasoner Adaptation Across Spectra.}
Structure-encoder baselines (E3NN, GATGNN, CrystalFormer, MEGNet, M3GNet) primarily test whether improved crystal representations and inductive biases (equivariance, attention-based message passing, periodic/angle-aware graph Transformers, universal interatomic potentials) suffice for accurate spectrum regression after pooling; sequence-oriented baselines (Mat2Spec, DOSTransformer) more directly address the functional/sequence nature of eDOS, with architectures or objectives designed around predicting structured spectra across energy. For DeepSciReasoner, we use the same two-stage architecture across all four targets, with two target-specific modifications: (i) for XANES, we extend the encoder with a learned embedding over the absorber atomic number (see \Cref{app:data}) so that the same crystal graph can produce different spectra for different absorbing elements; and (ii) on the phDOS, XANES, and Raman benchmarks we ensemble five independently-trained models (different seeds) at inference time by averaging their sum-normalized predictions, as is common practice for spectrum-prediction benchmarks~\citep{Kong2022}. All baselines are retrained from scratch on each target with the same train/val/test split and the same loss as our model's first stage; only the encoder architecture and the corresponding default optimizer settings differ across baselines.

\subsection{Evaluation Metrics}\label{app:eval}
\paragraph{Evaluation Metrics.}
We evaluate performance using task-specific metrics tailored to (i) full-spectrum prediction of eDOS, phDOS, XANES, and Raman, and (ii) valence-band (VB) gap identification.

\begin{itemize}
\item \textbf{Full-spectrum prediction} (spectrum regression).
Let \(p(\epsilon)\) denote the predicted normalized spectrum and \(p^\star(\epsilon)\) the DFT ground truth, discretized on \(N\) bins \(\{\epsilon_i\}_{i=1}^N\) along a spectral axis (energy for eDOS and XANES, wavenumber/frequency for phDOS and Raman) with values \(p_i\) and \(p_i^\star\). For eDOS and phDOS we report all four of mean absolute error (MAE), mean squared error (MSE), coefficient of determination (\(R^2\)), and Wasserstein distance (WD); for XANES and Raman, where bin-wise differences are dominated by sparse edge/peak structure, we report only \(R^2\) and WD to focus evaluation on shape and distributional fidelity.
\begin{itemize}
    \item {Mean Absolute Error (MAE):} \(\mathrm{MAE} = \tfrac{1}{N} \sum_{i=1}^N |p_i - p_i^\star|\).
    \item {Mean Squared Error (MSE):} \(\mathrm{MSE} = \tfrac{1}{N} \sum_{i=1}^N (p_i - p_i^\star)^2\).
    \item {Coefficient of determination (\(R^2\)):} \(R^2 = 1 - \sum_i (p_i - p_i^\star)^2 / \sum_i (p_i^\star - \bar{p}^\star)^2\), where \(\bar{p}^\star\) denotes the mean of the ground-truth spectrum.
    \item {Wasserstein distance (WD):} to capture distributional and shape differences between spectra, we compute the 1D Wasserstein-1 distance using cumulative distributions \(F(\epsilon) = \int_{-\infty}^{\epsilon} p(u)\,du\) and \(F^\star(\epsilon)\), \(\mathrm{WD} = \int |F(\epsilon) - F^\star(\epsilon)|\, d\epsilon\), implemented on the discrete spectral grid.
\end{itemize}

\item \textbf{VB gap identification} (band-edge classification).
To evaluate whether a model correctly identifies the valence-band gap, we cast the task as a binary
classification problem over energy bins, where bins within the ground-truth VB-gap region are labeled
positive and all others negative. We report
\begin{itemize}
    \item {Precision:} \(\mathrm{Precision} = \mathrm{TP}/(\mathrm{TP} + \mathrm{FP})\).
    \item {Recall:} \(\mathrm{Recall} = \mathrm{TP}/(\mathrm{TP} + \mathrm{FN})\).
    \item {F1 score:} \(\mathrm{F1} = 2 \cdot \mathrm{Precision} \cdot \mathrm{Recall} / (\mathrm{Precision} + \mathrm{Recall})\).
    \item {Matthews correlation coefficient (MCC):} 
    $$\mathrm{MCC} = (\mathrm{TP}\cdot\mathrm{TN} - \mathrm{FP}\cdot\mathrm{FN}) / \sqrt{(\mathrm{TP}+\mathrm{FP})(\mathrm{TP}+\mathrm{FN})(\mathrm{TN}+\mathrm{FP})(\mathrm{TN}+\mathrm{FN})},$$
    which is robust to strong class imbalance typical of gap regions.
\end{itemize}
\end{itemize}

\subsection{Computational Cost}\label{app:comp_cost}
We assess the computational efficiency of our proposed framework on a server equipped with 4 NVIDIA H100 GPUs and Intel Xeon Platinum 8462Y+. The implementation is based on PyTorch and PyTorch Geometric libraries. \Cref{tab:comp_cost} summarizes the computational budget for both Structure-Aware Spectrum Decoding and Mass-Conserving Iterative Refinement.

\begin{table}[ht]
\centering
\caption{Computational cost and hardware specifications on CalTech and TRI datasets.}
\label{tab:comp_cost}
\begin{small}
    \begin{sc}
        \begin{tabular}{p{4.5cm}cc}
        \toprule
        & \multicolumn{2}{c}{Dataset} \\
        \cmidrule(lr){2-3}
        Metric & CalTech & TRI \\
        \midrule
        \multicolumn{3}{l}{\textbf{Hardware Specifications}} \\
        \midrule
        GPU Specification & \multicolumn{2}{c}{1 $\times$ NVIDIA H100 (80GB)} \\
        CPU Specification & \multicolumn{2}{c}{Intel Xeon Platinum 8462Y+} \\
        \midrule
        \multicolumn{3}{l}{\textbf{Stage I: Encoder + Decoder Training}} \\
        \midrule
        Encoding Time / Crystal (ms) & 4.39 & 4.75 \\
        Decoding Time / Crystal (ms) & 8.64 & 10.93 \\
        Train Time / Epoch (m) & 1.43 & 0.82 \\  
        \midrule
        \multicolumn{3}{l}{\textbf{Stage II: MCIR Training}} \\
        \midrule
        Train Time / Epoch (m) & 1.46 & 0.85 \\
        \bottomrule
        \end{tabular}
    \end{sc}
\end{small}
\end{table}

\section{Additional Experimental Results}
\label{app:res}

\subsection{Data Scaling Analysis on TRI Dataset}\label{app:sca_tri}
\begin{figure}[ht]
    \centering
    \includegraphics[width=1\linewidth]{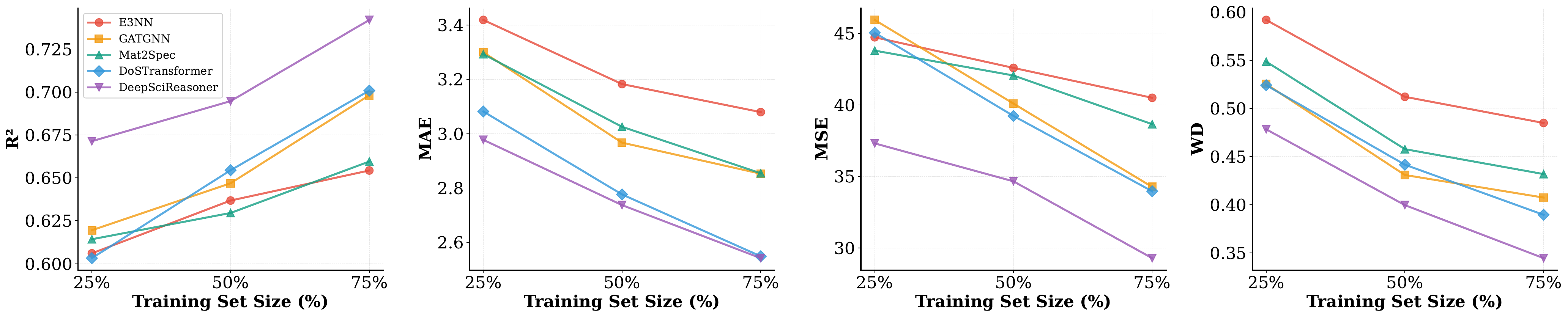}
    \caption{Data scaling behavior on the TRI dataset. Performance of different methods as a function of training set size (25\%, 50\%, 75\%) on R$^2$, MAE, MSE, and WD.} \label{fig:sca_trl}
\end{figure}

\Cref{fig:sca_trl} reports the scaling behavior on the TRI dataset as the size of training set increases from
25\% to 75\%. Trends closely mirror those observed on CalTech (\Cref{fig:sca}): across MAE, MSE,
\(R^2\), and WD, DeepSciReasoner consistently achieves the best performance in every data regime.
In the low-data (25\%) setting, DeepSciReasoner already establishes a clear margin over the strongest
baseline, highlighting the strong sample efficiency that stems from combining structure-aware decoding
with constraint-preserving refinement rather than relying on purely data-driven regression.
As the amount of training data increases, DeepSciReasoner improves more rapidly than competing
methods, with particularly pronounced gains in MSE and \(R^2\), indicating increasingly accurate
and variance-explaining spectral predictions.
Importantly, the performance gap does not diminish with additional data; instead, it remains stable
or widens across all metrics. This behavior suggests that DeepSciReasoner scales favorably with
dataset size and is able to convert increased supervision into improved spectral fidelity
(including distributional alignment measured by WD), whereas baseline models exhibit shallower
improvement trends and remain consistently inferior across the full data range.

\subsection{Visualization of Predicted eDOS Spectrum}\label{app:viz}
\begin{figure}[ht]
    \vspace{-5pt}
    \centering
    \includegraphics[width=1\linewidth]{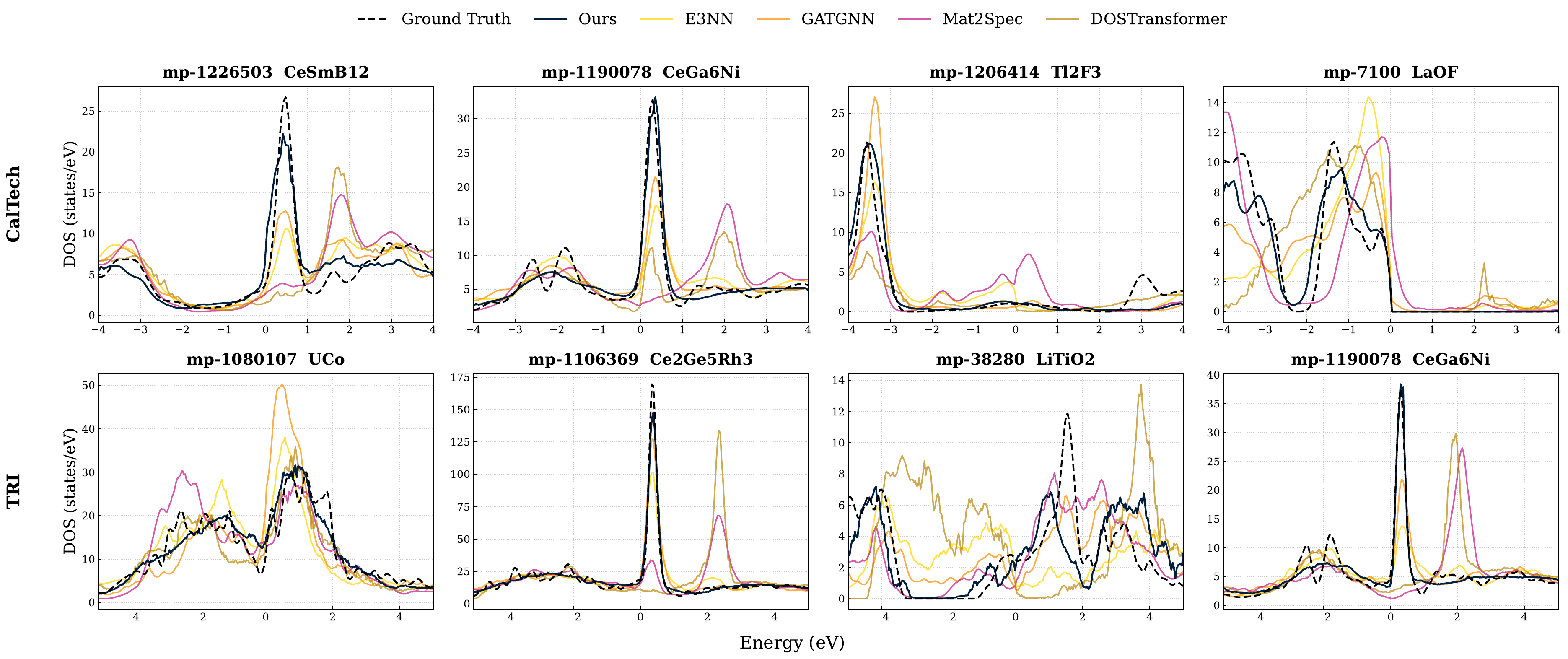}
    \caption{Representative examples from the CalTech (Top) and TRI (Bottom) test sets compare predicted eDOS of each method (including DeepSciReasoner and the baselines) against DFT-derived targets.} \label{fig:viz}
    \vspace{-5pt}
\end{figure}

\begin{figure}[h!]
    \vspace{-5pt}
    \centering
    \includegraphics[width=1\linewidth]{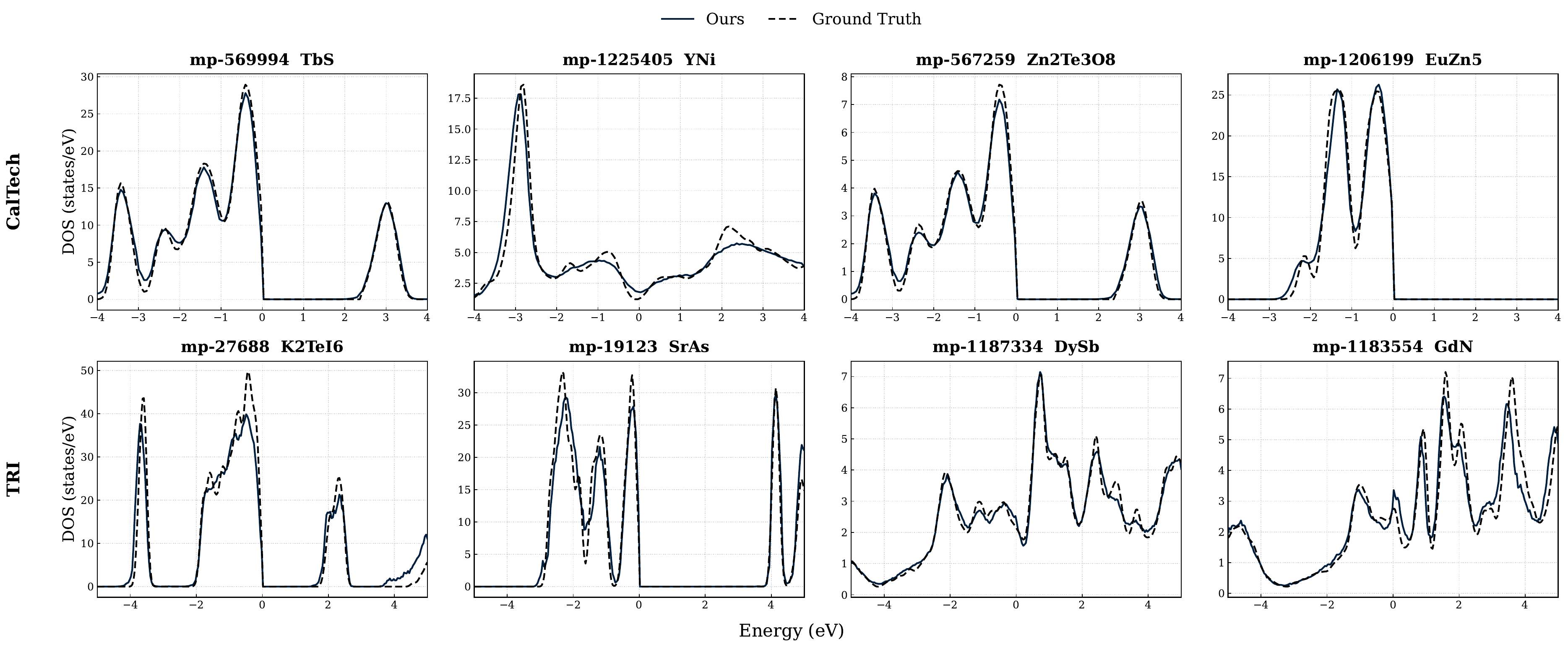}
    \caption{Representative examples from the CalTech (Top) and TRI (Bottom) test sets compare predicted eDOS of DeepSciReasoner against DFT-derived targets.} \label{fig:viz_ours}
\end{figure}

\Cref{fig:viz} provides qualitative evidence that complements the aggregate metrics by directly comparing predicted eDOS curves to the DFT-derived reference across representative CalTech (top row) and TRI (bottom row) examples. In these plots, our predictions consistently better reproduce two physically salient aspects of the spectrum: (i) the band edges/band gap region and (ii) sharp van Hove singularities. To make the comparison between our predictions and the ground truth clearer, we present several plots containing only these results in \Cref{fig:viz_ours}.

First, in materials with a pronounced gap or steep band onsets, DeepSciReasoner more faithfully captures the location and sharpness of the valence- and conduction-band edges, yielding a cleaner low-DOS region between them. Visually, this appears as a closer match to the ground-truth curve near the onset/offset points, with less spurious “leakage” of density into the gap compared with competing methods that either smear the edges or introduce artificial shoulders in the forbidden region. 

Second, across both datasets, our curves better preserve the positions and relative magnitudes of prominent peaks—the van Hove singularities that encode critical-point structure in the band dispersion. In several panels, baseline models either broaden these peaks (over-smoothing), shift them along the energy axis (misalignment), or create extra oscillatory peaks (overfitting artifacts). In contrast, our predictions more often place the dominant singularities at the correct energies and maintain their sharp, high-contrast profiles, leading to a closer qualitative match to the DFT lineshape in peak-dominated regions.

\end{document}